\documentclass[aps, pra, reprint, showpacs, superscriptaddress]{revtex4-1}

\usepackage{amsmath}
\usepackage{txfonts}
\usepackage{paralist}
\usepackage[usenames, dvipsnames, svgnames, table]{xcolor}
\usepackage{verbatim}
\usepackage{graphicx}
\usepackage[colorlinks=true, citecolor=blue, 
	        linkcolor=red, urlcolor=Green]{hyperref}

\usepackage[format=plain, justification=centerlast, small]{caption}
\usepackage{subcaption}

\newcommand{\emphb}[1]{\textbf{\textsf{#1}}}
\newcommand{\tss}{\textsuperscript}
\newcommand{\checkme}[1]{{\color{orange} #1}}
\newcommand{\textsub}[1]{{\text{\textsc{#1}}}}
\renewcommand{\textmu}{$\muup$}

\DeclareMathOperator{\real}{Re}

\begin{document}

\title{Broadband Measurement of Coating Thermal Noise in Rigid Fabry--P\'{e}rot Cavities}

\author{Tara Chalermsongsak}
\affiliation{LIGO Laboratory, California Institute of Technology, MS 100--36, Pasadena, CA 91125, USA}
\author{Frank Seifert}
\affiliation{Joint Quantum Institute, National Institute of Standards and Technology and University of Maryland, 100 Bureau Drive, Gaithersburg, MD 20899, USA}
\author{Evan D. Hall}
\author{Koji Arai}
\author{Eric K. Gustafson}
\author{Rana X Adhikari}
\affiliation{LIGO Laboratory, California Institute of Technology, MS 100--36, Pasadena, CA 91125, USA}

\begin{abstract}
We report on the relative length fluctuation of two fixed-spacer Fabry--P\'erot cavities with mirrors fabricated from silica/tantala dielectric coatings on fused silica substrates.
By locking a laser to each cavity and reading out the beat note $\hat{\nu} = \nu_1 - \nu_2$ of the  transmitted beams, we find that, for frequencies from 10\,Hz to 1\,kHz, the amplitude spectral density of beat note fluctuation is $\sqrt{S_{\hat{\nu}}(f)} = (0.5 \text{ Hz})/f^{1/2}$.
By careful budgeting of noise sources contributing to the beat note, we find that our measurement is consistent with the fluctuation in this band being dominated by the Brownian noise of the mirror coatings.
Fitting for the coating loss angle $\phi_\text{c}$, we find it equal to $4\times10^{-4}$.
We then use a Bayesian analysis to combine our measurement with previous observations, and thereby extract estimates for the individual loss angles of silica and tantala.
The testbed described in this article can be used in the future to measure the length noise of cavities formed with novel mirror coating materials and geometries.
\end{abstract}

\maketitle

\section{Introduction}

Thermal noise is an important fundamental noise source in precision experiments.

In the field of gravitational wave (GW) detection, thermal noise affects instruments such as Advanced LIGO, a large-scale Michelson interferometer with Fabry--P\'erot arm cavities 4\,km in length. Advanced LIGO will attempt to measure GW-induced spacetime fluctuations with a sensitivity of $1.4\times10^{-20}\text{ m/Hz}^{1/2}$ in the most sensitive band, around 200--500\,Hz.
It is predicted that this sensitivity will be limited in part by thermal noise in the high-reflectivity coating of the mirrors~\cite{Harry2010}.

Many groups have developed mathematical models to calculate coating thermal noise~\cite{Nakagawa2002, Harry2002, Somiya2009, Hong2013}.
However, due to these coatings' multilayer structure and uncertainties in the thin film material parameters (e.g., Young's moduli, Poisson ratios, and mechanical loss angles), thermal noise in coatings has not yet been thoroughly understood.
For this reason, an experiment which can measure coating thermal noise with high signal-to-noise ratio across a wide frequency band is necessary for a comprehensive verification of their performance.
 
Previously, direct measurements of thermal noise have been carried out with free-space cavities formed from large, suspended mirrors (e.g., Numata et~al.~\cite{Numata2003} and Black et~al.~\cite{Black2004_1}).
The nature of these suspensions is such that thermal noise can be observed only above a few hundred hertz; seismic motion becomes a limiting noise source at frequencies below 100\,Hz.

On the other hand, in the field of optical frequency metrology, a fixed spacer Fabry--P\'erot cavity is typically used as a stable reference for laser frequency.
By designing the shape of the spacer, and searching for vibration-insensitive support points, several groups have demonstrated that the total displacement noise of a rigid cavity can be very close to the thermal noise limit at frequencies around 0.01--1\,Hz~\cite{Ludlow2007, Alnis2008, Webster2007b}.
However, none have reported Brownian thermal noise in the frequency band relevant to ground based GW detectors. 

These motivations have led us to develop an experiment that uses fixed-spacer cavities to directly 
observe thermal noise in mirror coatings from 10\,Hz to 1\,kHz.
We demonstrate a method that can be used to measure thermal noise in SiO$_2$/Ta$_2$O$_5$ quarter-wavelength (QWL) coatings over two decades in frequency.

\section{Theory of Thermal Noise}
\label{sec:theory}

In this section we describe the fluctuation-dissipation theorem and its use in calculating thermal noise.

\subsection{Fluctuation-Dissipation Theorem}

Analysis of thermal noise begins with the fluctuation-dissipation theorem (FDT)~\cite{Callen1952}, which states that the more heavily damped a system is when driven by an external force, the noisier it is when sitting in its quiescent state.
The single-sided PSD of the system's generalized displacement $x(t)$ is given by
    \begin{equation}
        S_x(f) = \frac{k_\text{B} T}{\pi^2 f^2} \bigl|\real[Y(f)]\bigr|,
    \label{eq:fdt}
    \end{equation}
where $Y = 1 / Z$ is the mechanical admittance.
We define the system's mechanical impedance $Z$ as the complex frequency-domain response $F(f)/\dot{x}(f)$, where $F$ is the generalized force conjugate to $x$~\cite{Saulson1990}.

In considering the Brownian noise of a LIGO mirror, Saulson~\cite{Saulson1990} found an expression for $S_x(f)$ by computing $\bigl|\real[Y(f)]\bigr|$ separately for each of the normal modes contributing to the strain of the mirror.
However, this method is computationally expensive, and the result is not guaranteed to converge \cite{Levin1998}.
Instead of using modal expansion, one can use the so-called ``direct approach'' to compute $\bigl|\real[Y(f)]\bigr|$.
This was introduced by Gonz\'{a}lez and Saulson~\cite{Gonzalez1994} for computing thermal noise in suspensions, and was later applied to a laser mirror by Levin~\cite{Levin1998}.
In this approach, one calculates the thermal noise by applying a cyclic force, which causes power dissipation in a lossy system.
With the FDT, the dissipated power $W_\text{diss}$ and the PSD $S_x$ are related by
\begin{equation}
    \label{eq:FDT2}
    S_x(f) = \frac{2 k_B T}{\pi^2 f^2} \frac{W_\text{diss}}{F_0^2},
\end{equation}  
where $F_0$ is the magnitude of the applied force used to calculate the dissipated power.
In the case of a mirror whose position is interrogated by a laser beam, the cyclic ``force'' applied is a pressure with the same profile as the intensity of the beam.

\subsection{Types of Thermal Noise}

There are two known sources of thermal noise present in extended solid systems: mechanical loss and thermal dissipation.
Mechanical loss is responsible for \emph{Brownian noise}.
Thermal dissipation leads to temperature fluctuation, which in an optical system is converted to position fluctuation via the optic's coefficient of thermal expansion (CTE) $\alpha = (1/L)\partial L/\partial T$ and its thermorefractive coefficient $\beta = \partial n/\partial T$. 
The noise of this position fluctuation is called \emph{thermo-optic noise}.

\subsubsection{Brownian noise}

Mechanical loss arises from the microscopic structure of a material, such as impurities or dislocations.
It is represented by introducing an imaginary part to the Young's modulus of the material: $E = E_0(1+i\phi)$.
The quantity $\phi$ is referred to as the \emph{loss angle}, and in general may have a frequency dependence.
When a sinusoidal force is applied to a system with mechanical loss, the dissipated power due to the applied force is
\begin{equation}
    \label{eq:elasticU} 
    W_\text{diss} = 2\pi f U_0 \phi,
\end{equation}
where $U_0$ is the maximum energy of elastic deformation~\cite{Levin1998}.
If one is interested only in frequencies $f$ below the first mechanical resonance frequency of the system (as is the case with our reference cavities), it is sufficient to compute the stored energy $U_0$ in the presence of a static force.
The problem of evaluating $W_\text{diss}$ then reduces to a single elastostatics computation, which can be carried out using finite-element analysis (FEA) if necessary.
Together with eq.~\ref{eq:FDT2}, one can then calculate the Brownian contribution to the apparent position fluctuation of the mirror as sensed by a laser beam interrogating the mirror surface.

\subsubsection{Thermo-optic noise}

In contrast to Brownian noise, thermo-optic noise is related to thermal, rather than mechanical, dissipation; it arises from fluctuation in the temperature field $T(\mathbf{r},t)$ throughout the mirror~\cite{Zener1938}. 
To compute thermo-optic noise using the direct approach, one can apply either an imaginary force~\cite{Liu2000, Somiya2009} or imaginary heat~\cite{Levin2008, Evans2008} to the mirror's surface; the results will be the same if the stress inside the coating is uniform~\cite{Somiya2009}.
The applied force will cause temperature gradients inside the mirror through the equation of static stress balance.
Then, the temperature perturbation evolves according to the thermal diffusion equation (see, e.g., the treatment by Liu and Thorne~\cite{Liu2000} or Cerdonio et~al.~\cite{Cerdonio2001}).
Finally, the power dissipation due to the heat flow caused by the temperature gradient is given by the expression \cite[eq.~35.1]{L&L}
\begin{equation}
\label{eq:LL}
W_\text{diss} = \left\langle T\frac{dS}{dt} \right\rangle = \left\langle \int \frac{\kappa}{T} (\boldsymbol{\nabla} \delta T)^2 \mathrm{d}^3 r \right\rangle.
\end{equation}
Here $T$ is the unperturbed temperature of the system and $\delta T$ is the temperature perturbation due to the applied force $F_0$.
The entropy $S$ of the system changes due to the heat flux $-\kappa \boldsymbol{\nabla}(\delta T)$, and $\langle \cdots \rangle$ denotes an average over the period of oscillation of the force.
By substituting eq.~\ref{eq:LL} into eq.~\ref{eq:FDT2}, we can obtain the temperature fluctuation on the mirror sensed by a Gaussian laser beam.
This fluctuation couples into the electromagnetic response of the mirror via the CTE and $\partial n/\partial T$.

In the literature, the term ``thermoelastic noise'' refers to the effect from the change in position of the mirror surface due to thermal expansion of a substate and coating \cite{BGV1999, Liu2000, Cerdonio2001, Fejer2004}. On the other hand, ``thermorefractive noise'' refers to the phase fluctuation of the beam as it propagates through or reflects off the mirror, and it is a combined effect of both the CTE and $\partial n/\partial T$. 

For a Fabry--P\'{e}rot cavity with mirrors fabricated from multilayer dielectric coatings, thermorefractive noise in the substrate is much smaller than that in the coating~\cite{Heinert2011}: the beam passes through each substrate only once, but it reflects off the multilayer coating multiple times as it circulates inside the cavity.
Thus, for our experiment, we take thermorefractive noise into account only in the coating. Since both thermoelastic and thermorefractive noises have a common origin, they are computed in a coherent fashion and the combined effect is called thermo-optic noise~\cite{Evans2008}. For substrates and spacers, only thermoelastic noise will be considered. 

\section{Noise budget for fixed-spacer Fabry--P\'erot cavities}
\label{sec:noise_budget}

\begin{table}[tbp]
    \centering
    \begin{tabular}{l l c c}
        \toprule
        Symb.               & Description                  & Initial cav.    & Short cav.  \\
        \colrule
        $L$                 & Nominal spacer length        & 20.3~cm         & 3.68(3)~cm\footnote{Machining specification was $L = 1.45 \pm 0.01$ inches.}   \\
        $R_\text{sp}$       & Outer spacer radius          & 25.4~mm\footnote{\label{fn:longspacer}LIGO internal document D980670.}  & 19.0~mm  \\
        $r_\text{sp}$       & Inner spacer radius          & 6.4~mm\tss{\ref{fn:longspacer}}    & 5.1~mm  \\
        $R_\text{s}$        & Mirror substrate radius      & \multicolumn{2}{c}{12.7~mm} \\
        $\mathcal{R}$       & Mirror ROC\footnote{Uncertainty taken as 0.5\% of the nominal.}                   & \multicolumn{2}{c}{500(3) mm}  \\
        $\lambda$           & Laser wavelength             & \multicolumn{2}{c}{1064~nm}  \\
        $w$                 & Spot size on mirrors\footnote{Defined as the radius for which the intensity has fallen by $1/\mathrm{e}^2$ relative to the maximum intensity. Computed as $w = (\lambda\mathcal{R}/\pi)^{1/2}/(2\mathcal{R}/L-1)^{1/4}$.}    & 290~{\textmu}m  & 182.0(4)~{\textmu}m  \\
        $\mathcal{F}$       & Finesse                      & \multicolumn{2}{c}{$10\,000$} \\
        $\mathcal{T}$       & Power transmission (per mirror)   & \multicolumn{2}{c}{300~ppm}     \\
        $T$                 & Cavity temperature           & \multicolumn{2}{c}{306(1)~K} \\
        \colrule
        $E_\text{s}$        & Substrate Young modulus\footnote{The quantities $E_\text{sp}$, $\sigma_\text{sp}$, etc., for the spacer are taken to the identical to the quantities for the substrate.}  & \multicolumn{2}{c}{72(1)~GPa} \\
        $\sigma_\text{s}$   & Substrate Poisson ratio           & \multicolumn{2}{c}{0.170(5)} \\
        $\phi_\text{s}$     & Substrate loss angle              & \multicolumn{2}{c}{$1\times10^{-7}$} \\
        $\kappa_\text{s}$   & Subst. therm. conduct.            & \multicolumn{2}{c}{1.38~W/(m~K)} \\
        $C_\text{s}$        & Substrate heat capacity           & \multicolumn{2}{c}{$1.6\times10^6$~J/(K~m$^3$)} \\
        $\alpha_\text{s}$   & Substrate CTE                     & \multicolumn{2}{c}{$5.1\times10^{-7}$~K$^{-1}$}  \\ 
        \colrule
        $E_\text{L}$         & Young modulus of silica          & \multicolumn{2}{c}{72(1)\,GPa} \\
        $E_\text{H}$         & Young modulus of tantala\footnote{Nominal value and uncertainty from Crooks et~al.~\cite[tab.~6]{Crooks2006}.}         & \multicolumn{2}{c}{144(42)\,GPa} \\
        $n_\text{L}$         & Silica index of refraction\footnote{\label{fn:index_ref}Values from Evans et~al.~\cite[tab.~II]{Evans2008}.}   & \multicolumn{2}{c}{$1.45(1)$}  \\
        $n_\text{H}$         & Tantala index of refraction\tss{\ref{fn:index_ref}}  & \multicolumn{2}{c}{$2.06(1)$}  \\
        $N$                     & Number of coating layers\footnote{The first 27 layers are quarter-wavelength, and the top layer is a half-wavelength silica cap.}       & \multicolumn{2}{c}{28}        \\
        $d$                     & Coat. total thickness\footnote{Calculated as $d = 14 \lambda/4n_\text{Ta$_2$O$_5$} + (13+2)\lambda/4n_\text{SiO$_2$}$.}        & \multicolumn{2}{c}{4.53(7)~{\textmu}m} \\
        \botrule
    \end{tabular}
    \caption{Parameters for test cavities.}
    \label{tab:cavity_params}
\end{table}

In this section we present the assumptions and formulas used to generate the thermal noise contributions to the noise budget.
Numerical values of the relevant parameters and symbols are given in table \ref{tab:cavity_params}.

\subsection{Mirror substrate noise}

\subsubsection{Substrate Brownian noise}

Levin~\cite[eq.~2]{Levin1998} computed the Brownian noise for a mirror substrate in the limit that the spot size $w$ is much smaller than the radius $R_\text{s}$ of the mirror:
    \begin{equation}
        S_x^{(\text{subBr})}(f) = \frac{2 k_\text{B} T}{\pi^{3/2} f}
            \frac{\bigl(1-\sigma_\text{s}^2\bigr)\phi_\text{s}}{w E_\text{s}}.
        \label{eq:substrate_brownian}
    \end{equation}
The spot size is defined as the $1/\mathrm{e}^2$ falloff in intensity.
$E_\text{s}$, $\sigma_\text{s}$, and $\phi_\text{s}$ are, respectively, the Young modulus, Poisson ratio, and loss angle of the substrate.
Later, Bondu et~al.~\cite[eq.~14]{Bondu1998} computed corrections to the above formula for the case when $w$ is not much smaller than $R_\text{s}$, but we have found that these corrections are not necessary for our system.

\subsubsection{Substrate thermoelastic noise}

The thermoelastic noise for a mirror substrate was computed by Braginsky et~al.~\cite{BGV1999} for the case of a half-infinite substrate in the adiabatic limit $\ell_\text{th} \ll w$, where $\ell_\text{th} = \sqrt{\kappa_\text{s}/(2 \pi C_\text{s} f)}$ is the thermal diffusion length at frequency $f$, and $\kappa_\text{s}$ and $C_\text{s}$ are, respectively, the thermal conductivity and the heat capacity per unit volume of the substrate.
Non-adiabatic corrections for low frequencies and small beam sizes were computed by Cerdonio et~al.~\cite[eq.~20]{Cerdonio2001}:
    \begin{equation}
        S_x^{(\text{subTE})}(f) = \frac{4 k_\text{B} T^2}{\sqrt{\pi}}
            \frac{\alpha_\text{s}^2 \bigl(1+\sigma_\text{s}\bigr)^2 w}{\kappa_\text{s}}
            J\bigl(f/f_\text{T}\bigr),
        \label{eq:substrate_TE}
    \end{equation}
where  $f_\text{T} = \kappa_\text{s}/\pi w^2 C_\text{s}$, and $J(f/f_\text{T})$ is a non-elementary function whose asymptotes are $2/\bigl(3\sqrt{\pi f/f_T}\bigr)$ for $f/f_\text{T} \ll 1$ and $1/\bigl(f/f_T\bigr)^2$ for $f/f_\text{T} \gg 1$; the full expression is
    \begin{equation}
        J(f/f_\text{T}) = \left(\frac{2}{\pi}\right)^{1/2} \int\limits_0^\infty \mathrm{d}u \int\limits_{-\infty}^\infty \mathrm{d}v \; \frac{u^3 \mathrm{e}^{-u^2/2}}{(u^2 + v^2)\bigl[(u^2 + v^2)^2 + (f/f_\text{T})^2\bigr]}.
        \label{eq:Jint}
    \end{equation}

\subsection{Noise in mirror coatings}

\subsubsection{Coating Brownian noise}

The Brownian thermal noise contribution of a thin film on a half-infinite substrate can be expressed as~\cite{Nakagawa2002} 
\begin{equation}
    \label{eq:Nakagawa_BR_coat}
    S_x^{\text{(cBR)}}(f) = \frac{4 k_B T}{\pi^2 f} \frac{(1 + \sigma_s)(1 - 2\sigma_s)}{E_s}
        \frac{d}{w^2} \phi_{c},
\end{equation}
where $d$ is the total thickness of the coating, and $\phi_\text{c}$ is the coating's loss angle.

This equation assumes that the elastic properties of substrate and the thin coating are the same, and that all the coating properties are isotropic.
Due to the multilayer structure of the amorphous materials, the coating loss and elastic properties may be anisotropic.
For this reason, authors such as Harry et~al.~\cite{Harry2002} decompose coating loss and elastic deformation into parallel ($\parallel$) and perpendicular ($\perp$) directions relative to the mirror normal.
Then, in accordance with eq.~\ref{eq:elasticU}, the total dissipated energy can be written as $W_\text{diss} = 2\pi f (U_{\perp} \phi_{\perp} + U_{\parallel} \phi_{\parallel})$. 

However, as argued by Hong et al.~\cite{Hong2013}, $\phi_{\perp}$ and $\phi_{\parallel}$ are not a suitable choice to be consistently used as the loss angles of a material, since the corresponding energies $U_{\perp}$ and $U_{\parallel}$ can sometimes be negative.
Instead, $W_\text{diss}$ should be decomposed into bulk (``B'') and shear (``S'') contributions: $W_\text{diss} = 2\pi f( U_\text{B} \phi_\text{B} + U_\text{S} \phi_\text{S})$.

For SiO$_2$/Ta$_2$O$_5$ coatings, the individual loss angles (either $\phi_{\perp}$ and $\phi_{\parallel}$, or $\phi_\text{B}$ and $\phi_\text{S}$) are not well known, and knowledge of the individual material properties is also limited.
These uncertainties will propagate forward toward the estimate of the loss angle~\cite{Hong2013}.

In this work, we assume the equality of $\phi_\text{B}$ and $\phi_\text{S}$, but we stress that there is no fundamental reason to assume this, nor indeed is there reason to assume equality of the elastic parameters of the substrate and the coating.
Nevertheless, if we assume that the coating is described by a single loss angle $\phi_\text{c}$, and that the elastic properties of the coating and substrate are similar, then the results of Harry et~al.~\cite{Harry2002} and Hong et~al.~\cite{Hong2013} reduce to eq.~\ref{eq:Nakagawa_BR_coat}.
The ``coating loss angle'' $\phi_\text{c}$ as defined in equation~\ref{eq:Nakagawa_BR_coat} should be viewed not as a physical parameter, but as a figure of merit which is related to the various loss angles and material parameters of each coating material.
%


\subsubsection{Coating thermo-optic noise}

An expression for thermo-optic noise in coatings is given by Evans et~al.~\cite[eq.~4]{Evans2008}:
    \begin{equation}
        S_x^{(\text{cTO})}(f) = S_T(f) \, \Gamma_\text{tc} \,
            \left[\bar{\alpha}_\text{c} d - \bar{\beta}\lambda - \bar{\alpha}_\text{s} d C_\text{c} /C_\text{s}\right]^2.
    \label{eq:coating_TO}
    \end{equation}
Here $S_T(f)$ is the temperature fluctuation of a bare substrate as sensed by an interrogating beam. In the adiabatic regime, it is given by~\cite{Levin2008}
    \begin{equation}
        S_T(f) = \frac{2 k_\text{B} T^2}{\pi^{3/2} w^2 \sqrt{\kappa_\text{s} C_s f}}.
    \label{eq:TO_spectrum1}
    \end{equation}
$\Gamma_\text{tc}$ is a correction for $S_T(f)$ in the presence of a coating layer.
The term in brackets in eq.~\ref{eq:coating_TO} determines how temperature flucutation $S_T$ is converted into displacement fluctuation $S_x$.
$\bar{\alpha}_\text{c}$, $\bar{\beta}$, and $C_\text{c}$ are the effective thermal expansion coefficient, effective thermorefractive coefficient, and heat capacity per unit volume of the coating.
The quantities $\bar{\alpha}_\text{s}$ and $C_\text{s}$ are the thermal expansion coefficient and heat capacity per unit volume of the substrate.
The complete formalism for computing the various thermal expansion and thermorefractive coefficients is summarized by Evans et~al.~\cite[appx.~A~and~B]{Evans2008}.

%

Similar to substrate thermoelastic noise, the temperature fluctuation in eq.~\ref{eq:TO_spectrum1} can be corrected for small beam size and low frequencies by extending the calculation by Braginsky et~al.~\cite{BGV2000}. The result is given by Martin~\cite[\S 3.3.2]{Mike2013}:
    \begin{equation}
        S_T(f) = \frac{2\sqrt{2} k_\text{B} T^2}{\pi \kappa_\text{s} w} M\bigl(f/f_\text{T}\bigr).
    \label{eq:TO_spectrum}
    \end{equation}
$M(f/f_T)$ is a non-elementary function whose asymptotes are $\sqrt{\pi/2}$ for $f/f_T \ll 1$ and $(2f/f_\text{T})^{-1/2}$ for $f/f_T \gg 1$.
The full expression is
    \begin{equation}
        M(f/f_\text{T}) = \real\left[ \int\limits_0^\infty \mathrm{d}u\; u\, \mathrm{e}^{-u^2/2} \sqrt{\frac{u^2 + i f / f_\text{T}}{u^4 + (f / f_\text{T})^2}}\right].
    \label{eq:Mint}
    \end{equation}

Note that $\Gamma_\text{tc}$ in Evans et~al.~\cite{Evans2008} is calculated assuming that $\ell_\text{th} \ll w$.
For SiO$_2$/Ta$_2$O$_5$ QWL coatings, $\ell_\text{th} = (44 \text{ \textmu m}) \times \sqrt{(100 \text{ Hz})/f}$, as calculated using the material parameters of silica and tantala, along with the formalism described by Evans et~al.~\cite{Evans2008}.
For a beam with spot size $w = 200 \text{ \textmu m}$, this correction factor should still be valid above 25~Hz.
However, a thorough calculation has yet to be done.

\begin{figure}[tbp]
    \centering
    \includegraphics[width = 3.5in]{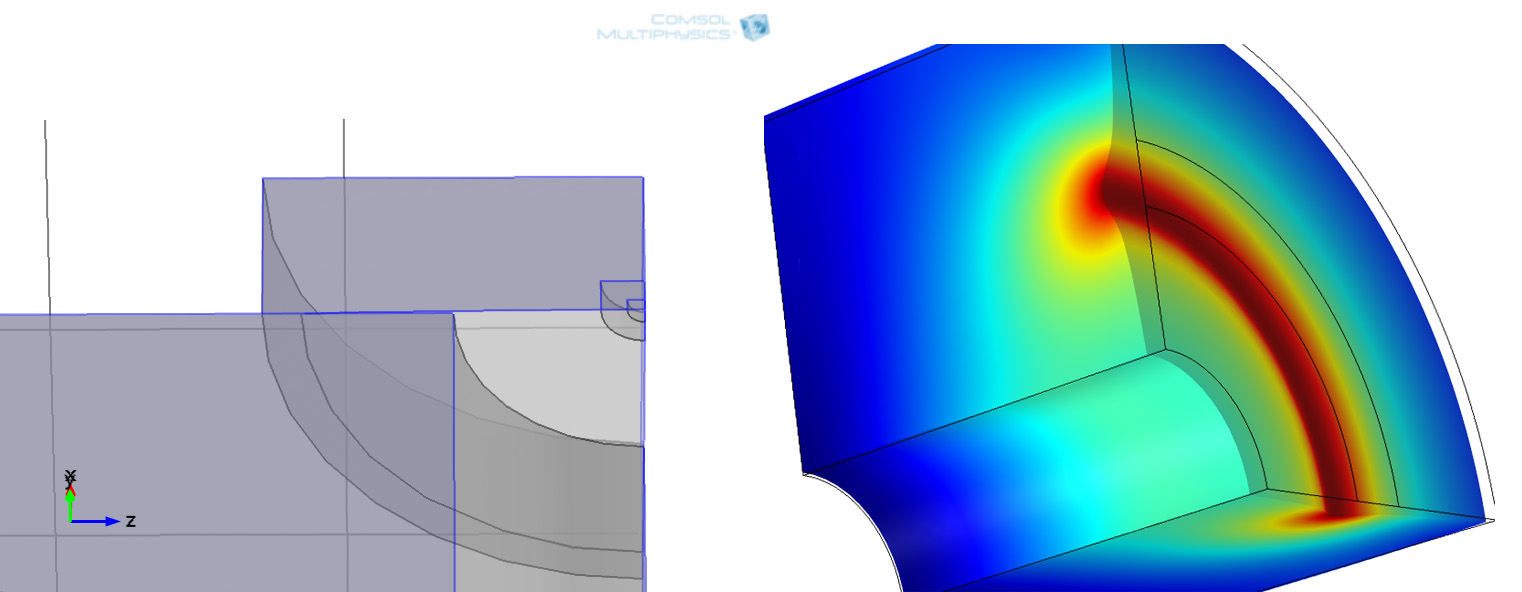}
    \caption{Image of FEA model used to predict thermal noise from the fused-silica cavity spacers.
    Left, 1/8 of the model is used with a symmetric boundary condition on three planes to reduce the computation time.
    Since most of the deformation will occur close to the applied force, to further minimize the calculation time, only the small volume at the center of the mirror has very fine mesh size while the mesh size is larger far away from the beam.
    Right, the deformation on the spacer due to the applied force on the mirror (not shown).
    This model can be used to calculate the elastic energy stored in the spacer.}
    \label{fig:comsolbr}
\end{figure}

\subsection{Noise in spacer}

\subsubsection{Spacer Brownian noise}

The length fluctuation due to Brownian noise in a cylindrically symmetric spacer of outer radius $R_\text{sp}$ and inner radius $r_\text{sp}$ was worked out by Kessler et~al.~\cite{Kessler2012a}, building on earlier work by Numata et~al.~\cite{Numata2004}:
    \begin{equation}
        S_x^{(\text{spBr})}(f) = \frac{4k_\text{B}T}{\pi f} \frac{L \phi_\text{sp}}{2\pi E_\text{sp} \bigl(R_\text{sp}^2 - r_\text{sp}^2\bigr)}.
    \label{eq:spacer_brownian}
    \end{equation}
However, this formula assumes that the outer radii of the mirror and the spacer are the same, and are fully contacted.
In general, the outer radius of the spacer is larger than the mirror radius, and only a thin annulus on the outer edge of the mirror is optically contacted to the spacer.
To estimate the Brownian noise more accurately, an FEA simulation along with the direct approach is used to calculate the stored elastic energy (see fig.~\ref{fig:comsolbr}).
Then, using eq.~\ref{eq:FDT2}, we obtain the displacement noise.
The power spectral density of the displacement noise computed from the FEA is about a factor of 2 larger than that of eq.~\ref{eq:spacer_brownian}.

\subsubsection{Spacer thermoelastic noise}

To estimate the level of thermoelastic noise in the spacer, we follow the method outlined by Liu and Thorne~\cite[eq.~13]{Liu2000}:
    \begin{equation}
        S_x^{(\text{spTE})}(f) = \frac{2 k_\text{B} T}{\pi^2 f^2} \kappa_\text{sp} T \left[\frac{E_\text{sp} \alpha_\text{sp}}{(1-2\sigma_\text{sp}) C_\text{sp}}\right]^2
            \int \frac{\bigl[\boldsymbol{\nabla} (\boldsymbol{\nabla}\cdot\mathbf{u}) \bigr]^2}{F_0^2} \; \mathrm{d}^3 r,
    \label{eq:spacer_TE}
    \end{equation}
where $\mathbf{u}(\mathbf{r})$ is the displacement field of the spacer in response to a static pressure from a force $F_0$ applied to the mirror faces.
To evaluate the integral in eq.~\ref{eq:spacer_TE}, we use the same FEA model as described above for computing the spacer Brownian noise.
The calculation is performed under the adiabatic approximation, since the diffusion length $\ell_\text{th}$ is much smaller than the width of the contact area between the spacer and the mirror.
For an annulus with a thickness of 2\,mm, the assumption $\ell_\text{th} \ll w$ should be valid down to a few millihertz.
At very low frequencies, where the assumption on $\ell_\text{th}$ is not satisfied, the expected thermoelastic noise is smaller than the adiabatic prediction~\cite{Cerdonio2001}.   

\subsection{Photothermal Noise}
Fluctuation in laser power, either from shot noise or from classical intensity noise, induces a local temperature change in both coating and substrate.
Because of the thermal expansion and thermorefractive coefficients of the mirror substrate and the coating, the temperature gradient caused by the absorbed laser power couples into the cavity's displacement noise.
This is called photothermal noise.
As with thermo-optic noise, the effect in the substrate is mostly thermoelastic.
This noise source was first considered in a restricted regime by Braginsky et~al.~\cite{BGV1999}.
The full expression for photothermal noise in a mirror substrate, valid for small beam size and low frequencies, is~\cite{Cerdonio2001}
    \begin{equation}
    \label{eq:sub_photo}
        S_x^{\text{(PT)}}(f) = \frac{2}{\pi^2} \frac{(1+\sigma_\text{s})^2}{\kappa_\text{s}^2} \mathcal{S}_\text{abs} K(f/f_T).
    \end{equation}
where
    \begin{equation}
    \label{eq:sub_photo_K}
        K(f/f_T) = \left| \frac{1}{\pi} \int\limits_0^\infty \mathrm{d}u \int\limits_{-\infty}^\infty \mathrm{d}v\, \frac{u^2 \mathrm{e}^{-u^2 / 2}}{(u^2 + v^2) (u^2 + v^2 + if/f_\text{T})}\right|^2
    \end{equation}
and
    \begin{equation}
        \mathcal{S}_\text{abs} = \delta P(f) \frac{2\mathcal{F}/\pi}{ 1 + (f/f_\text{cav})^2} \chi_\text{abs}.
    \end{equation}
$\delta P(f)$ is the input power fluctuation, $\chi_\text{abs}$ is the absorption coefficient of the mirror, and $f_\text{cav} = f_\text{FSR} / (2\mathcal{F})$ is the cavity pole frequency.
 
The effects from the coating (both thermoelastic and thermorefractive) were later included in the work of Farsi et~al.~\cite[appendix]{Farsi2012}, who treat all the contributions from substrate and coating coherently.
We do not reproduce their formulas here.
The effect can be measured directly by modulating the power of the laser and observing the corresponding length change of the cavity.
   
Generally, relative intensity noise (RIN) in a laser is much higher than its shot noise limit and causes excessive photothermal noise.
This will be discussed in section \ref{sec:tech_noise}.

Here, $S_x^\text{(PT)}$ is the RIN-induced photothermal noise for a single mirror of a cavity.
The noise on the two mirrors is coherent, and so the total effect on the cavity is $4S_x^\text{(PT)}(f)$. 
We do not consider photothermal effects in the cavity spacer, since these effects only occur at frequencies below our measurement band.
 
\subsection{Total thermal noise in cavities}

Finally, we note that the length noise $S_L$ of a Fabry--P\'erot cavity involves the sum of the contributions from two mirrors and a single spacer:
    \begin{align}
        S_L &= 2 S_x^{(\text{cBr})} + 2 S_x^{(\text{cTO})} + 2 S_x^{(\text{subBr})} + 2 S_x^{(\text{subTE})} \nonumber \\
            &\hspace{2em} + S_x^{(\text{spBr})} + S_x^{(\text{spTE})} + 4S_x^{(\text{PT})}.
    \label{eq:cavity_length_noise}
    \end{align}
In the subsequent sections, we consider a number of technical and environmental noise sources which must be added to $S_L$ in order to arrive at the experimentally measured noise spectrum.


\section{Description of experiment}
\label{sec:experiment}

In this section we describe the testbed we have developed to measure the beat note fluctuation $S_{\hat{\nu}}(f)$ of our cavities.

\subsection{Cavity as a frequency reference}
\begin{figure}[tbp]
    \centering
    \includegraphics[width = 0.4\textwidth]{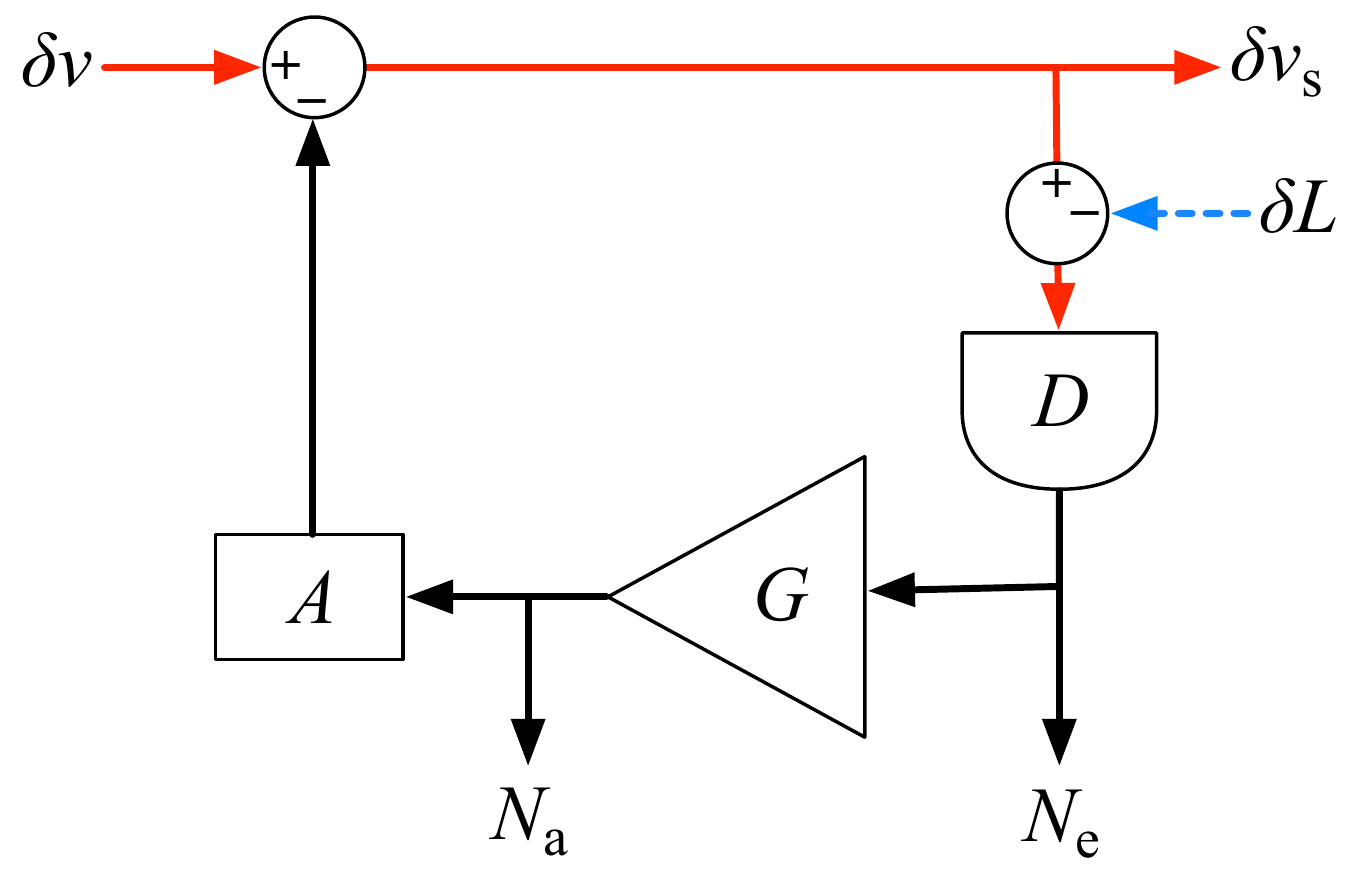}
    \caption{Block diagram of the PDH setup used for laser frequency locking.
    $\delta\nu$ is the free-running frequency noise of the laser, and $\delta\nu_\text{s}$ is the suppressed frequency noise, or the frequency noise of the transmitted beam behind the cavity.
    $\delta L$ is the cavity's length fluctuation, which is converted to frequency noise via the PDH lock.
    $D$ is frequency discriminator, which uses an rf photodiode and associated demodulation electronics to convert frequency fluctuation into an error signal with voltage $N_\text{e}$.
    $G$ is the electronic gain of the servo.
    $A$ is the actuator, which takes the control signal voltage $N_\text{a}$ and actuates on the laser frequency.
    The fact that $A$ is summed with a minus sign indicates that negative feedback is occurring.
    The minus sign from $\delta L$ means the displacement noise of the cavity is compared to the laser frequency.}
    \label{fig:PDHblock}
\end{figure}

Figure~\ref{fig:PDHblock} shows a block diagram of a laser that is frequency-locked to a reference cavity using Pound--Drever--Hall (PDH) locking~\cite{DH1983}.

The laser has free-running noise $\delta\nu$.
The frequency discriminator $D$, electronic servo gain $G$, and actuator response $A$ combine to produce the open-loop gain $H = DGA$.
When the loop is engaged, the suppressed frequency noise $\delta\nu_\text{s}$ of the laser becomes
\begin{subequations}
    \begin{align}
        \delta\nu_\text{s} &= \frac{\delta\nu}{1+H} + \frac{H}{1+H} \times \frac{c}{L\lambda} \delta L \\
        &\approx \frac{\delta\nu}{H} + \frac{c}{L\lambda} \delta L  \hspace{5em} \text{for } |H| \gg 1.
    \end{align}
    \label{eq:PDHnoise}
\end{subequations}
Within the loop bandwidth, where the magnitude $|H|$ of the open-loop gain is large, the displacement noise $\delta L$ of the cavity is impressed onto the frequency noise of the laser: $\delta\nu_\text{s} \approx (c/L\lambda) \delta L$.
The power spectral density of the frequency noise is given by $S_\nu(f) = |\delta\nu_\text{s}|^2$.  

To measure the frequency noise of the laser when locked to the cavity, we compare the transmitted beam with another transmitted beam from a similar cavity with an independent, frequency-stabilized laser.
Because of the slightly different lengths of the two cavities, the two beams have different frequencies, $\nu_1$ and $\nu_2$.
When directed onto an RF photodiode, the combined beam results in a beat note with frequency $\hat{\nu} = \nu_1 - \nu_2$.
The frequency noise of this beat note has a PSD $S_{\hat{\nu}} = S_{\nu_1} + S_{\nu_2}$. As described below, we read out this beat note using a phase-locked loop (PLL).




\subsection{Setup}

In this section, we describe two experimental setups used for observing coating thermal noise.
The first setup measures the noise from two 20.3~cm reference cavities.
The second setup, which is conceptually similar to the first one, measures coating thermal noise from two 3.68~cm reference cavities. 
\begin{figure}[tbp]
    \centering
    \includegraphics[width = 3.5in]{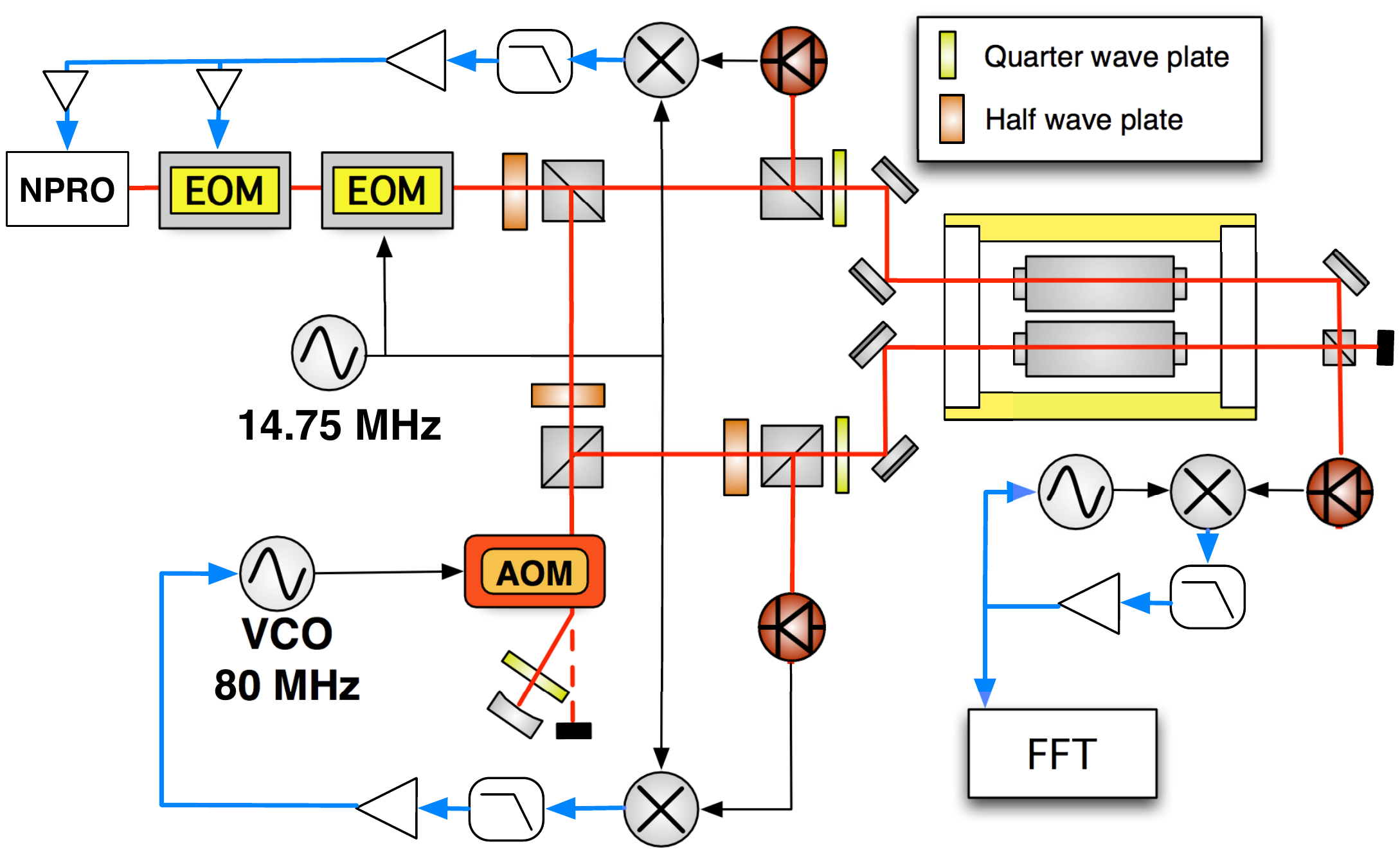}
    \caption{The prototype one-laser setup for measuring the coating thermal noise of LIGO reference cavities.
    An Nd:YAG laser is stabilized to one reference cavity.
    The second beam is split from the main beam and locked to the second cavity after being double-passed through an acousto-optic modulator (AOM).
    The transmitted beams are used to measure the length noise by measuring their beat signal.}
    \label{fig:setup1}
\end{figure}

\subsubsection{One-laser setup}

A diagram of the one-laser setup is shown in Figure~\ref{fig:setup1}.
In this setup, both interrogation beams were provided by a single Nd:YAG non-planar ring oscillator (NPRO) laser with a vacuum wavelength of $\lambda = 1064$~nm.
Approximately 1~mW of light was incident on each cavity, with visibility $\eta$ of more than 0.95.
The main beam was frequency-locked to one of the cavities by actuating on the NPRO crystal with a piezo-electric transducer (PZT), as well as actuating on a broadband electro-optic modulator (EOM) placed in the optical path.
For the second cavity, part of the laser beam was sent through a double-pass acousto-optic modulator (AOM) in order to frequency-shift the light before entering the cavity.
Frequency locking to the second cavity was achieved by actuating on the AOM.
Laser light was injected through the vacuum chamber windows and into the cavities, where it was kept resonant via the PDH locking technique.
The photodiodes and electronics used to implement the frequency stabilization were designed to achieve a loop with unity-gain frequency (UGF) of nearly 1\,MHz, and to have a noise floor below the frequency noise of the cavities.
The transmitted beams were recombined and directed onto an RF photodiode, producing an RF beat note measured with a PLL and a spectrum analyzer.

\subsubsection{Test cavities} 

The reference cavities are formed by optically contacting laser mirrors to cylindrical fused-silica spacers.
The mirror substrates are commercially available fused silica with a 25.4\,mm diameter and 6.4\,mm thickness, and with a 0.5\,m radius of curvature (ROC).
The coatings were fabricated by Research Electro-Optics via ion-beam sputtering.
They consist of 28 alternating layers of silica (SiO$_2$) and tantala (Ta$_2$O$_5$).
The first 27 layers are each deposited to a thickness of $\lambda/4n$, where $n$ is the refractive index of the layer material.
The final layer is silica, and in order to give the appropriate interference condition it is deposited to a thickness of $\lambda/2n$.
The transmission of each mirror is approximately 300~ppm.
Using these mirrors, we initially constructed two symmetric cavities using fused-silica spacers with length $L = 20.3$\,cm.
Both substrates and spacers are made of fused silica because of its low mechanical loss and small thermal expansion coefficients.  

Each cavity is fitted with a pair of O-rings close to the cavity's Airy points. 
Each cavity sits on a pair of teflon blocks with a semicircular cut, and each block has a transverse V-shaped groove to keep an O-ring in place.
The cavities are placed side by side on a double-stack seismic isolation platform.
The resonances of this platform all lie below 10\,Hz.
The cavities and the platform are housed inside a temperature-stabilized vacuum chamber with the pressure below $10^{-7}$ torr.
The use of a single platform and chamber endows the beat measurement with some amount of common-mode rejection of seismic and ambient temperature noise. The optical table for the setup sits on pneumatic legs which have a resonant frequency around 1 Hz.

\begin{figure*}[tbp]
    \centering
  \includegraphics[width=\textwidth]{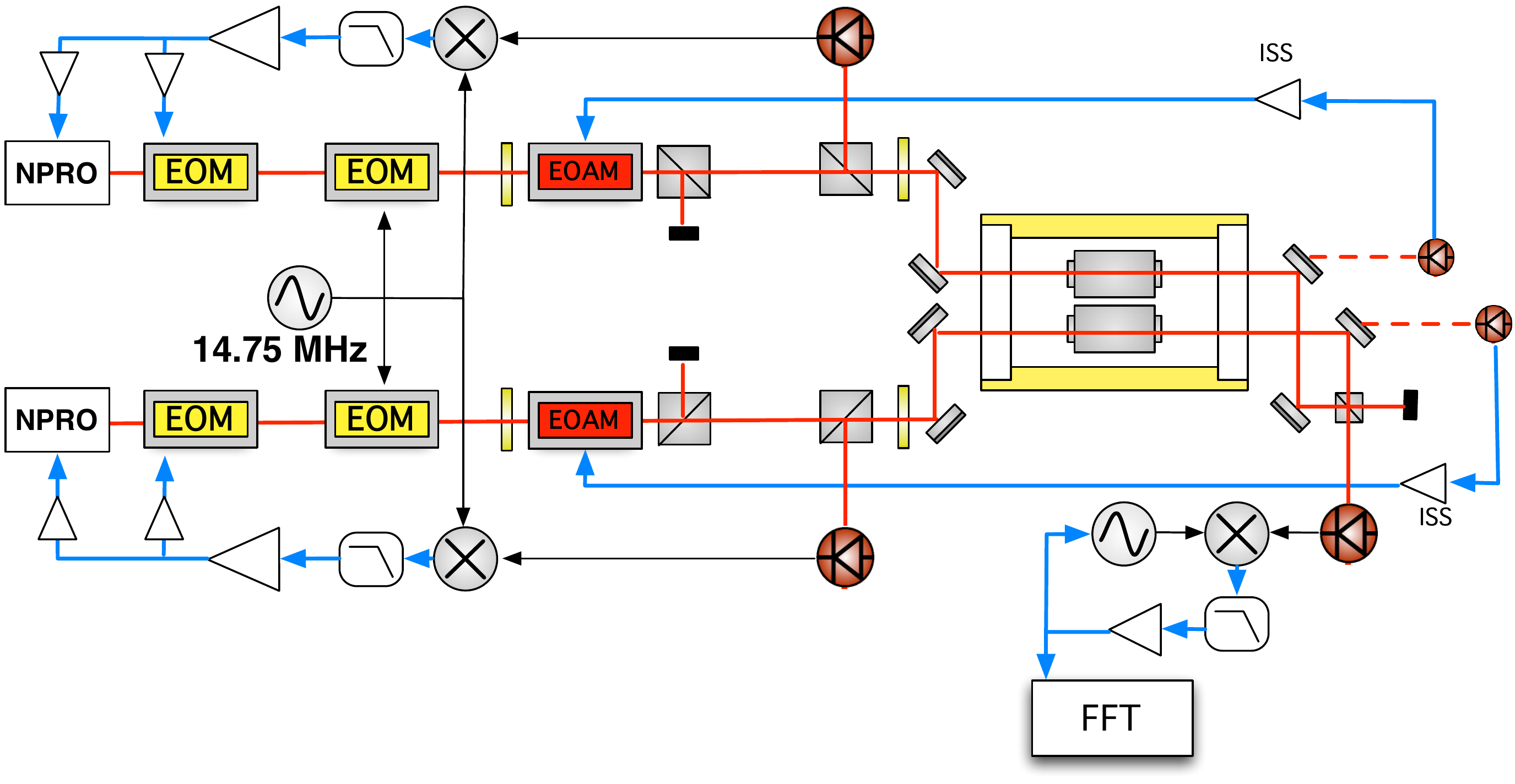}
    \caption{Setup for measuring the coating thermal noise using two independent lasers.
    Each laser is stabilized to one of the two identical cavities. 
    The readout scheme is the same as for the one-laser setup.}
    \label{fig:setup2}
\end{figure*}

\subsubsection{Two-laser setup} 

We found that the one-laser setup discussed above produced a measurement with a low signal-to-noise ratio.
Therefore, we subsequently constructed two shorter cavities using similar mirrors from the same coating run and developed a two-laser setup shown in Figure~\ref{fig:setup2}.
The use of shorter cavities increases the observed frequency noise, since $\delta\nu / \nu = \delta L / L$ for small cavity length fluctuations.
There were several considerations that placed a lower limit for the allowable length of the new cavities.
First, it should be possible to use a heater to tune each cavity length by half of a free spectral range, so that the beat note $\hat{\nu}$ can be brought within the bandwidth of the readout photodiode.
A cavity that is too short would require excessive heating in order to achieve this.
Additionally, the cavity must form a stable optical resonator.
Finally, the length must be chosen so that no low-order transverse laser modes resonate simultaneously with the TEM$_{00}$ mode.
With these considerations in mind, we chose a cavity length of 3.68\,cm.

In addition, compared to the previous, longer cavities, these cavities have a smaller spot size. The combined effects of shorter length and smaller spot size mean that the observed coating Brownian noise should increase by a factor of 9, in accordance with eq.~\ref{eq:Nakagawa_BR_coat}.

This setup is symmetric; the PDH error signal from each cavity is used to actuate on an independent NPRO and on a broadband EOM.
The use of two lasers also allows larger possible range for the beat frequency; in the previous setup, this was constrained by the operational range of the AOM.

For each path, 1\,mW of light is incident on each cavity.
The visibilities of both cavities exceed 0.9, indicating that the incident beams have a nearly 
Gaussian spatial mode and that the cavities are close to critically coupled.

In this setup, the relative intensity noise (RIN) in both cavities becomes uncorrelated, and so an intensity stabilizion system (ISS) is required.
In each path, an electro-optic amplitude modulator (EOAM) is used to suppress the laser's RIN, and thereby decrease the photothermal noise to below the estimated thermal noise of the coatings.



\subsubsection{Beat note frequency readout}

\begin{figure}[tbp]
    \centering
    \includegraphics[width=0.4\textwidth]{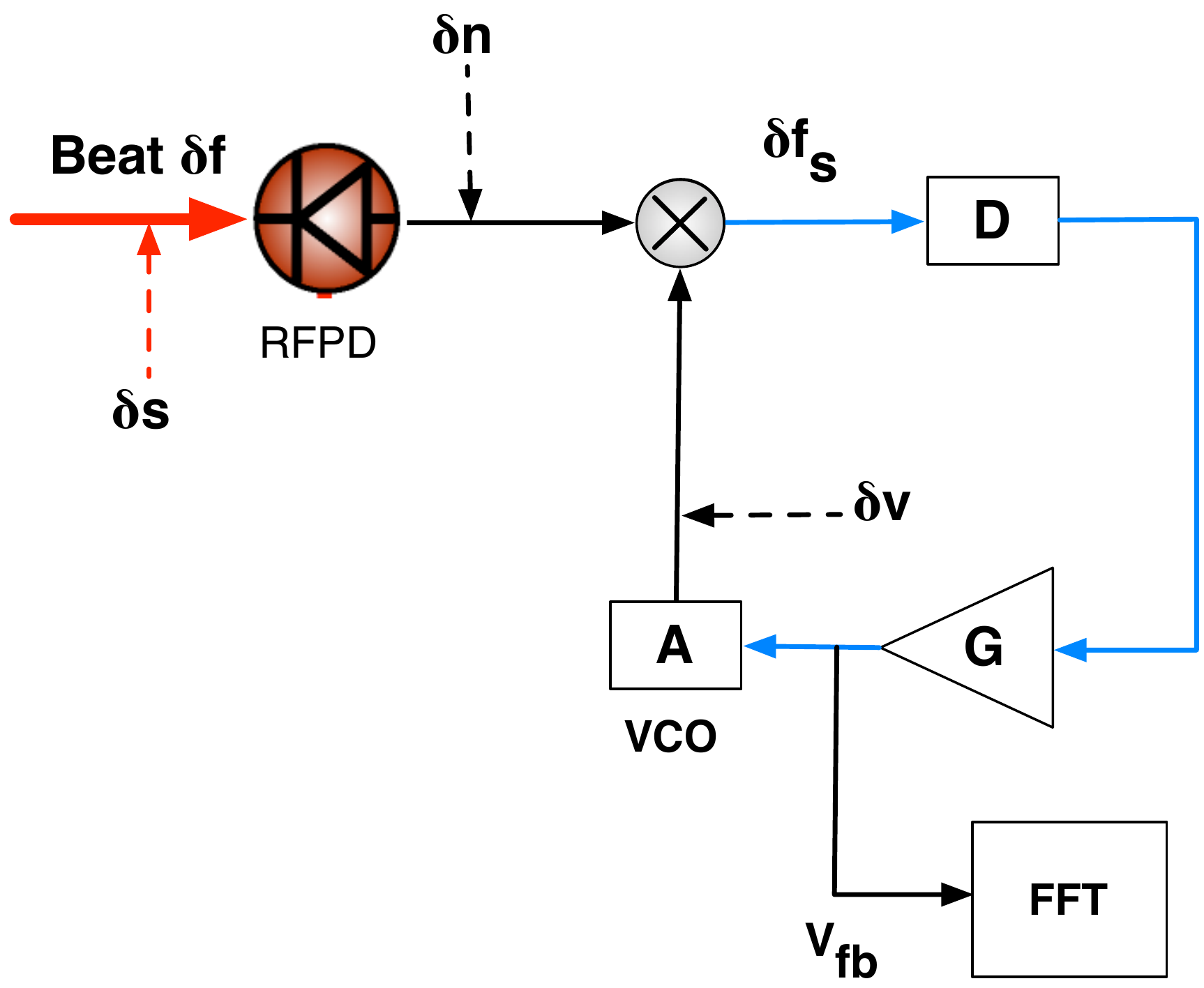}
    \caption{Block diagram of the phase-locked loop (PLL) used to read out the beat note fluctuation.
    The main noise sources associated with the PLL are photocurrent shot noise, $\delta s$; photodiode amplifier noise, $\delta n$; and VCO frequency noise, $\delta\nu$.
    Generally, $\delta s$ and $\delta n$ have flat spectral densities in terms of current and voltage, respectively.
    However, since the PLL is a phase detector whose output is then used to actuate on frequency, these noises contribute a frequency noise which rises with Fourier frequency $f$.}
    \label{fig:PLLblock}
\end{figure}

To read out the beat note frequency, we use a phase-locked loop (PLL).
A block diagram for the PLL is shown in fig.~\ref{fig:PLLblock}.

The two transmitted beams are directed onto a single RF photodiode, where they beat against each other to produce an RF signal at approximately 100\,MHz.
This signal is mixed with a voltage-controlled oscillator (VCO) of similar frequency and then low-passed at several megahertz, producing a baseband signal.
We then amplify this signal (labeled $V_\text{fb}$ in the diagram) and use it as a control signal to actuate on the VCO, thereby forming a phase-locked loop. 
This control signal gives a linear readout of the frequency noise of the beat note, which is the incoherent sum of the displacement noise from the two cavities.
The calibration to convert the voltage $V_\text{fb}$ to frequency fluctuation is measured by observing the output frequency of the VCO while varying the input voltage. 
The open loop gain of our PLL has a UGF of 50\,kHz.

\subsection{Technical and environmental noise sources}
\label{sec:tech_noise}
Both setups discussed in the previous section have similar technical and environmental noise sources.

\begin{figure}[tbp]
    \begin{subfigure}[tbp]{0.22\textwidth}
        \includegraphics[width=\textwidth]{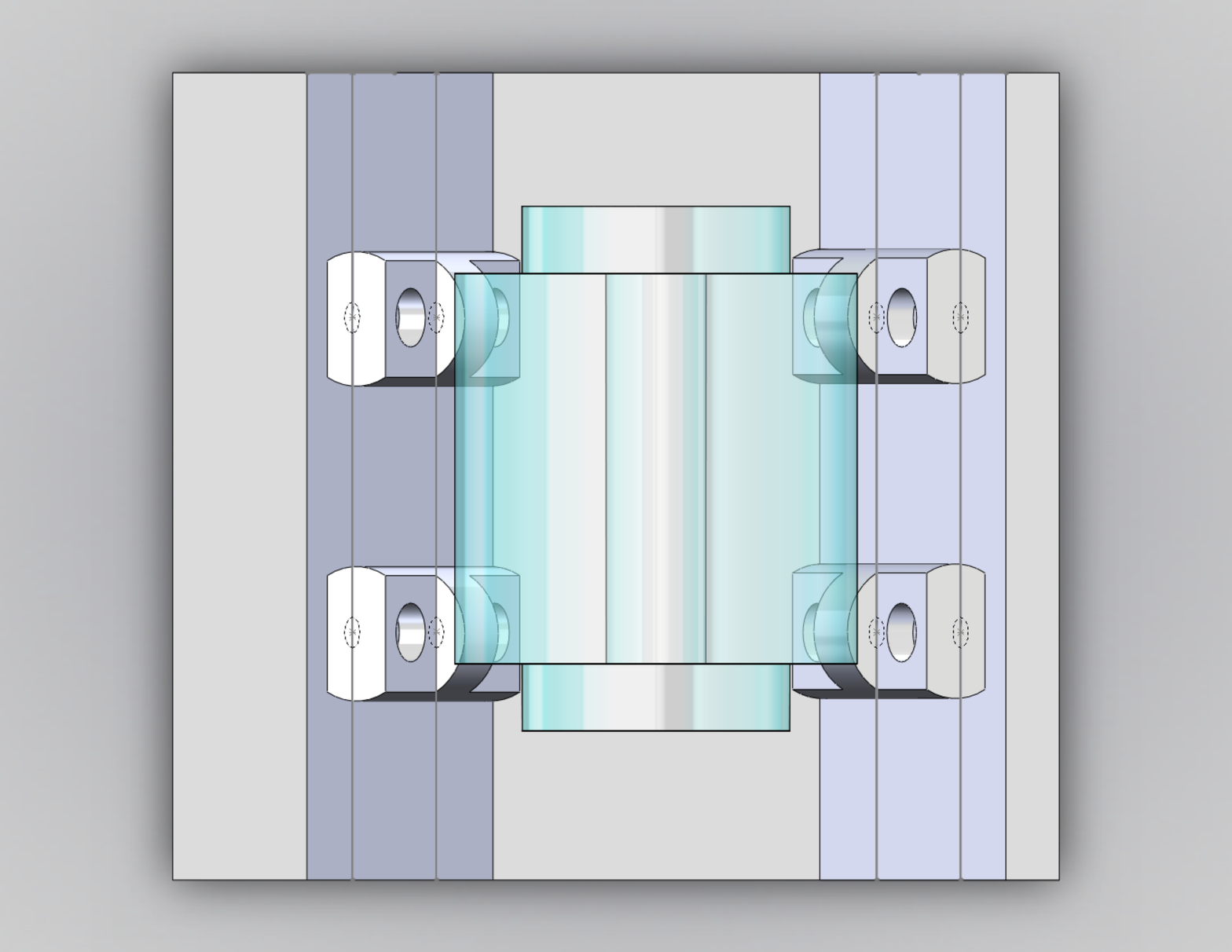}
        \caption{Top-down view}
        \label{fig:supports_top}
    \end{subfigure}
    \begin{subfigure}[tbp]{0.22\textwidth}
        \includegraphics[width=\textwidth]{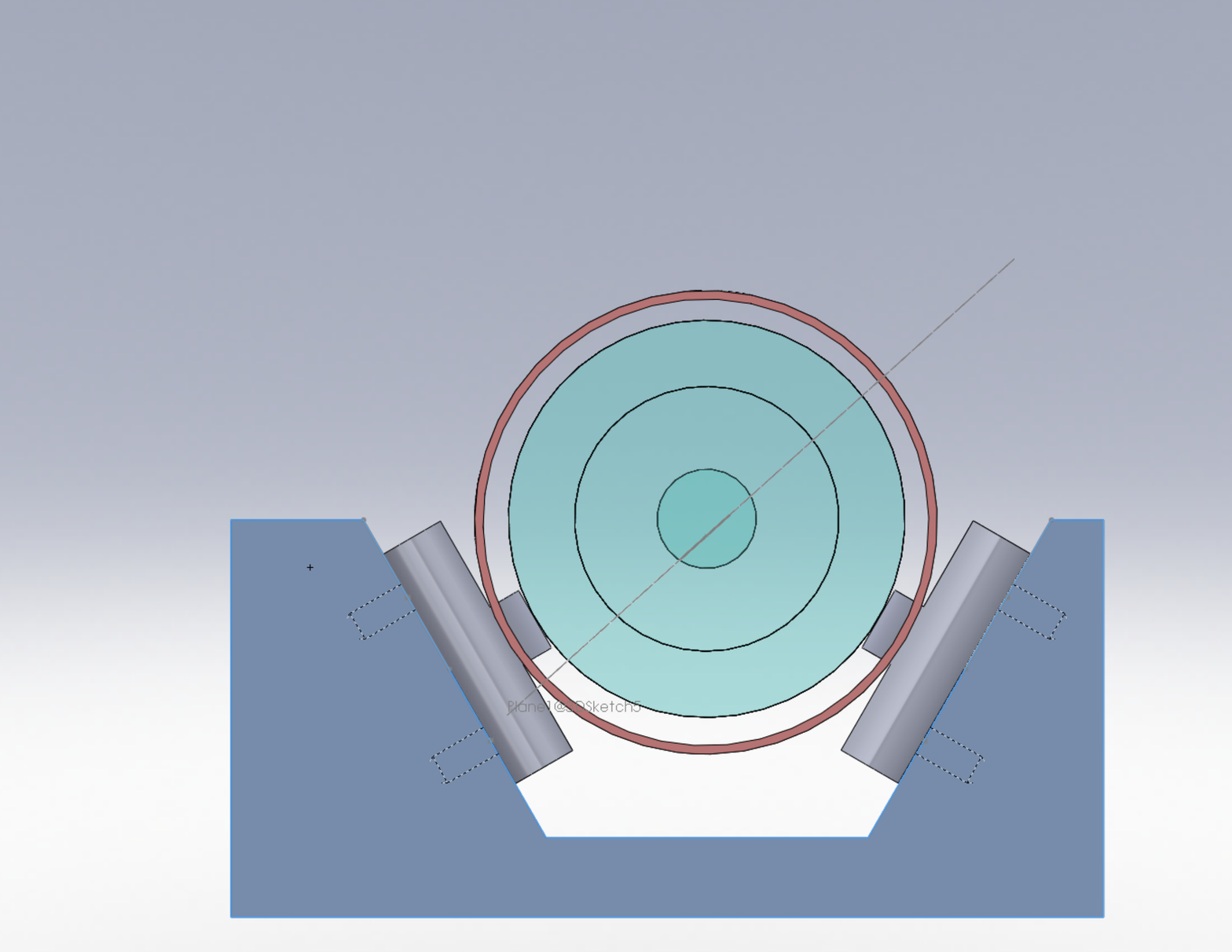}
        \caption{Axial view}
        \label{fig:supports_axial}
    \end{subfigure}
    \caption{Cavity mounting and supports for 3.7\,cm cavities.
    The locations of the four contacts were chosen for superior rejection of vertical seismic noise, as determined by FEA simulation.
    In the axial view, the red circle is the thermal shield used for temperature control.
    In the top-down view, this shield is not shown.}
    \label{fig:supports}
\end{figure}

\subsubsection{Seismic and vibrational noise}

Cavity bending due to vibration is known to cause significant displacement noise in a reference cavity.
To minimize this effect, some groups have explored different methods for supporting laser reference cavities---for example, by cutting or drilling support points into the spacer, or by holding the cavity vertically~\cite{Nazarova2006, Webster2007a, Millo2009}.
Based on our previous experience and FEA of the seismic coupling, we determined that the direct seismic coupling could be kept small enough with horizontal cavities with nodal supports.
  
In the two-laser setup, each cavity is mounted on four supports cut from cylindrical rods and placed orthogonally to the spacer to achieve approximately a point contact.
The support geometry is shown in Fig.~\ref{fig:supports}.
The rods are made from polyether ether ketone (PEEK) because of its compatibility with high vacuum.
The support positions were chosen based on ease of machining and on FEA of the susceptibility of the cavity to seismic noise.
At the chosen spot, if we take mounting errors ($\pm\,0.5$\,mm) and common mode rejection into account, the coupling from acceleration into cavity strain is estimated to be $6\times10^{-12}\,\text{ m}^{-1}\,\text{s}^2$. 

\subsubsection{PDH shot noise}

For each cavity, the ultimate lower limit to the laser's frequency noise suppression is set by the shot noise of the light falling on the RF photodiode when the cavity is on resonance. The PSD of this lower 
limit is~\cite{Fritschel:1998gr, Rana2004}
    \begin{equation}
        S_P^{(\text{PDHshot})}(f) = 2h\nu P_0 \bigl[J_0(\Gamma)^2(1-\eta) + 3J_1(\Gamma)^2\bigr],
        \label{eq:pdh_shot}
    \end{equation}
where $h$ is the Planck constant, $\Gamma$ is the phase modulation index ($\Gamma \approx 0.2$~rad for our system), $\eta$ is the visibility, and $J_0$ and $J_1$ are the zeroth and first Bessel functions of the first kind, respectively.

\subsubsection{Residual (RF) Amplitude Modulation}

The EOM used to perform the PDH modulation was temperature-stabilized with insulation and a heater, and then the polarization of the beam was adjusted to minimize any residual amplitude modulation (RAM), which can add a 
false offset to the PDH error signal 
(see, e.g., the discussion by Ishibashi et~al.~\cite{Hall2002}) 

\subsubsection{Photothermal Noise}

\begin{figure}[tbp]
    \centering
    \includegraphics[width = 3.7in]{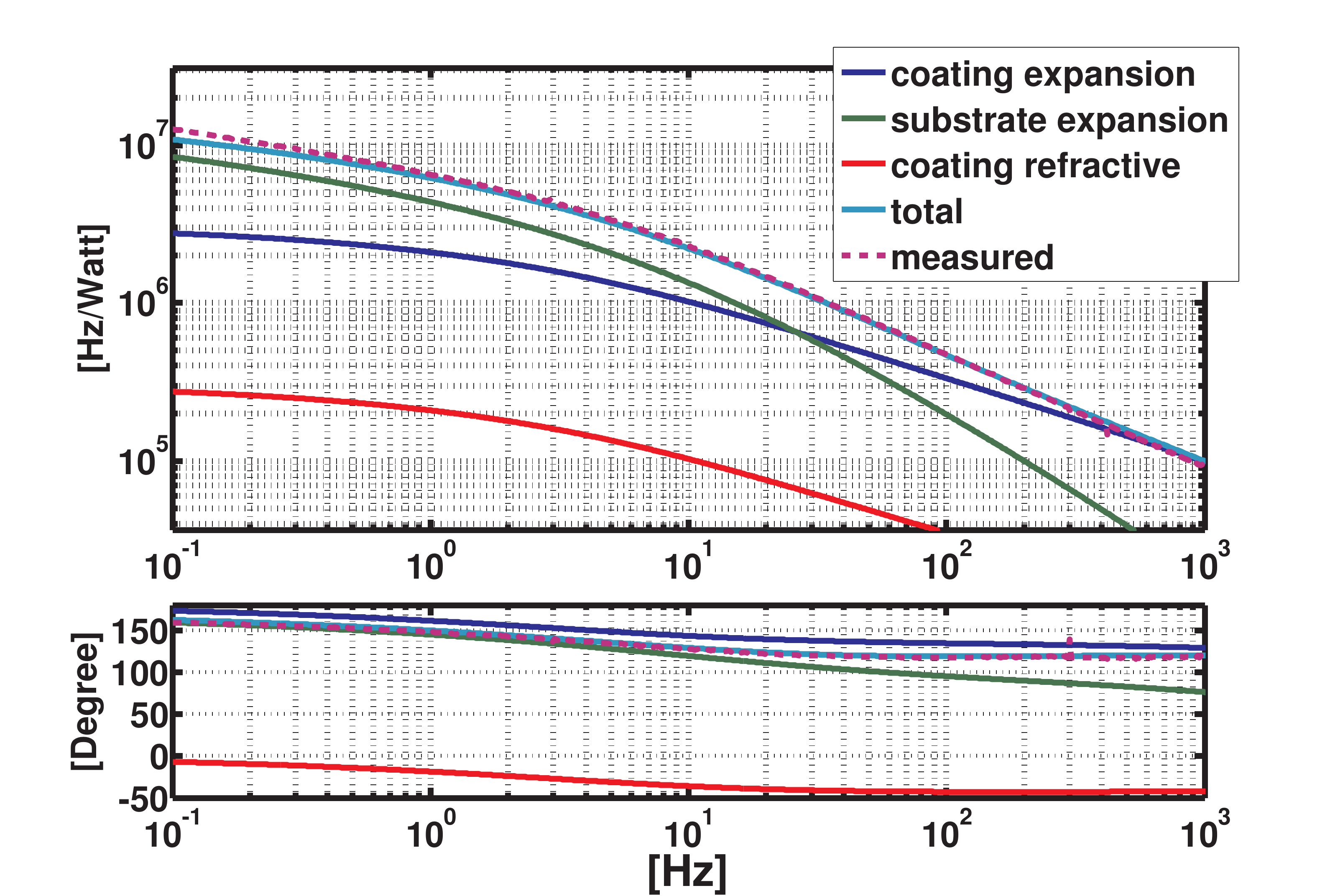}
    \caption{A swept sine measurement of beat note frequency 
    fluctuation in response to RIN-induced photothermal noise.
    Both the amplitude (top) and phase response (bottom) agree with the calculations from Farsi et~al.
    For our coatings, the greatest effects are thermal expansion from substrate and coating.}
    \label{fig:farsi}
\end{figure}

As discussed in section \ref{sec:noise_budget}, fluctuation in laser power changes the effective cavity length via the thermoelastic and thermorefractive coefficients.
In the case of a beam whose intensity fluctuation is shot-noise limited, the photothermal noise is negligible compared to Brownian thermal noise and thermoelastic noise~\cite{BGV1999}.
However, for a laser with significant intensity noise above the shot-noise limit, the photothermal noise can be much higher.
In the case of the one-laser setup, this excess photothermal effect appears in both cavities as a common-mode noise.
However, this is not the case for the two-laser setup, and so the photothermal effect has to be carefully characterized and factored 
into the noise budget.
By using the EOAM in each path to modulate the input power 
(see Fig.~\ref{fig:setup2}), we can observe the corresonding 
modulation in the beat note frequency using the PLL readout.
As shown in Figure~\ref{fig:farsi}, the results are comparable with the calculations given in Farsi, et~al.~\cite{Farsi2012} with the 
assumption of 5\,ppm absorption on each mirror.
Together with the measured RIN in the transmitted cavity beams, 
the estimated frequency noise due to RIN-induced photothermal 
noise can be added to the noise budget.

\subsubsection{PLL noise}

Noise sources add into the PLL at several points in the loop, as shown in Figure~\ref{fig:PLLblock}.
In the photodiode, there is shot noise from the photocurrent ($\delta s$) and electronic noise from the internal amplifier ($\delta n$). 
Additionally, there is frequency noise from the VCO ($\delta v$).
We have measured these noises and included them in the noise budget.

\section{Results}
\label{sec:results}

The measured PSD of the beat note frequency fluctuation $S_{\hat{\nu}}(f)$ is given by the sum of the cavity length noise $S_L(f)$ from both cavities, as well as the technical frequency noises:
    \begin{equation}
        S_{\hat{\nu}} = 2\left(\frac{c}{L\lambda}\right)^2 S_L(f) + S_{\hat{\nu}}^{(\text{tech})}(f)
        \label{eq:length_to_beat}
    \end{equation}
where $S_{\hat{\nu}}^{(\text{tech})}$ contains the contributions from the residual frequency noise, PLL readout noise, and seismic noise.

\begin{figure*}[tbp]
    \centering
    \includegraphics[width=0.85\textwidth]{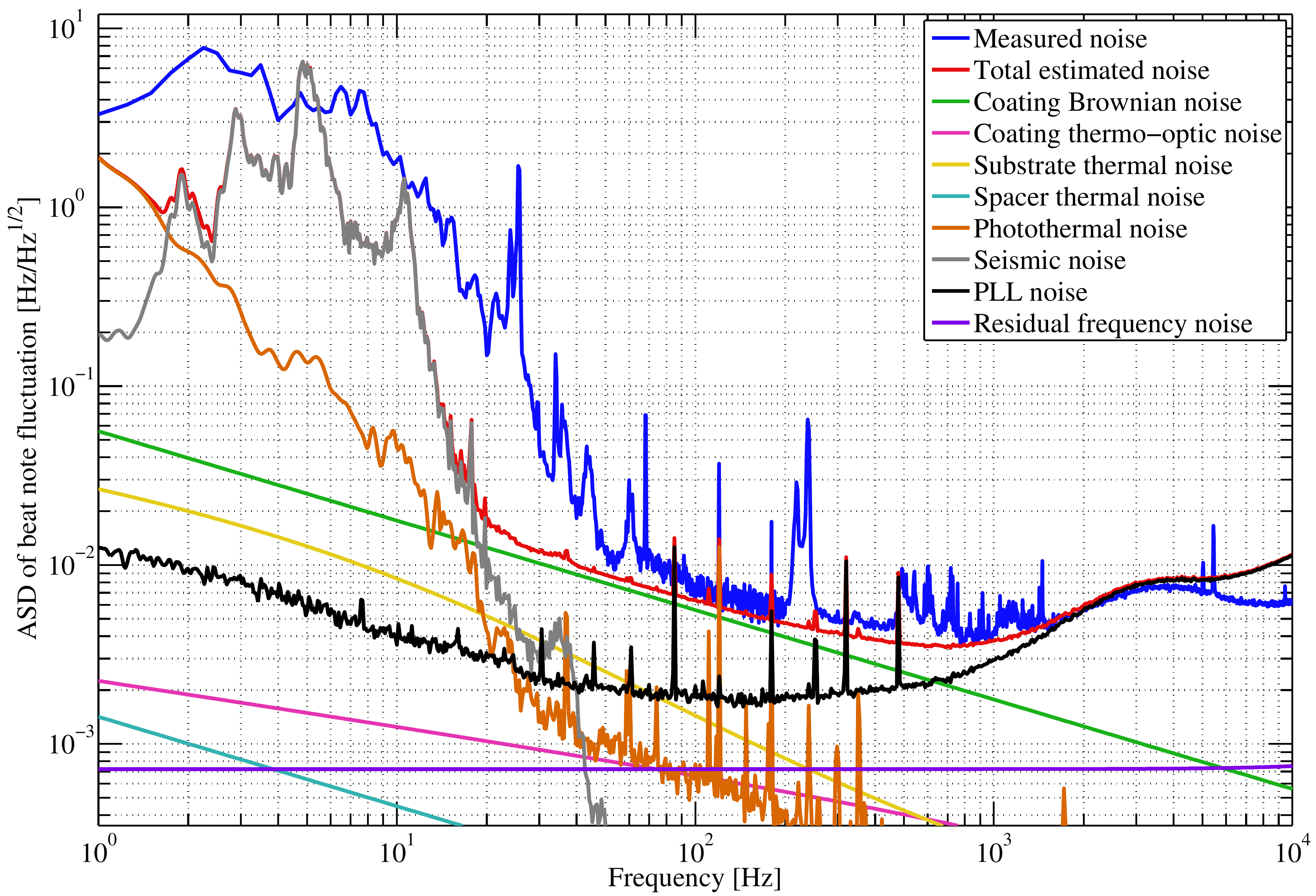}
    \caption{Amplitude spectral density $\sqrt{S_{\hat{\nu}}(f)}$ of beat note from 20.3~cm cavities using one-laser setup.}
    \label{fig:nb_beat_long}
\end{figure*}
\begin{figure*}[tbp]
    \centering
    \includegraphics[width=0.85\textwidth]{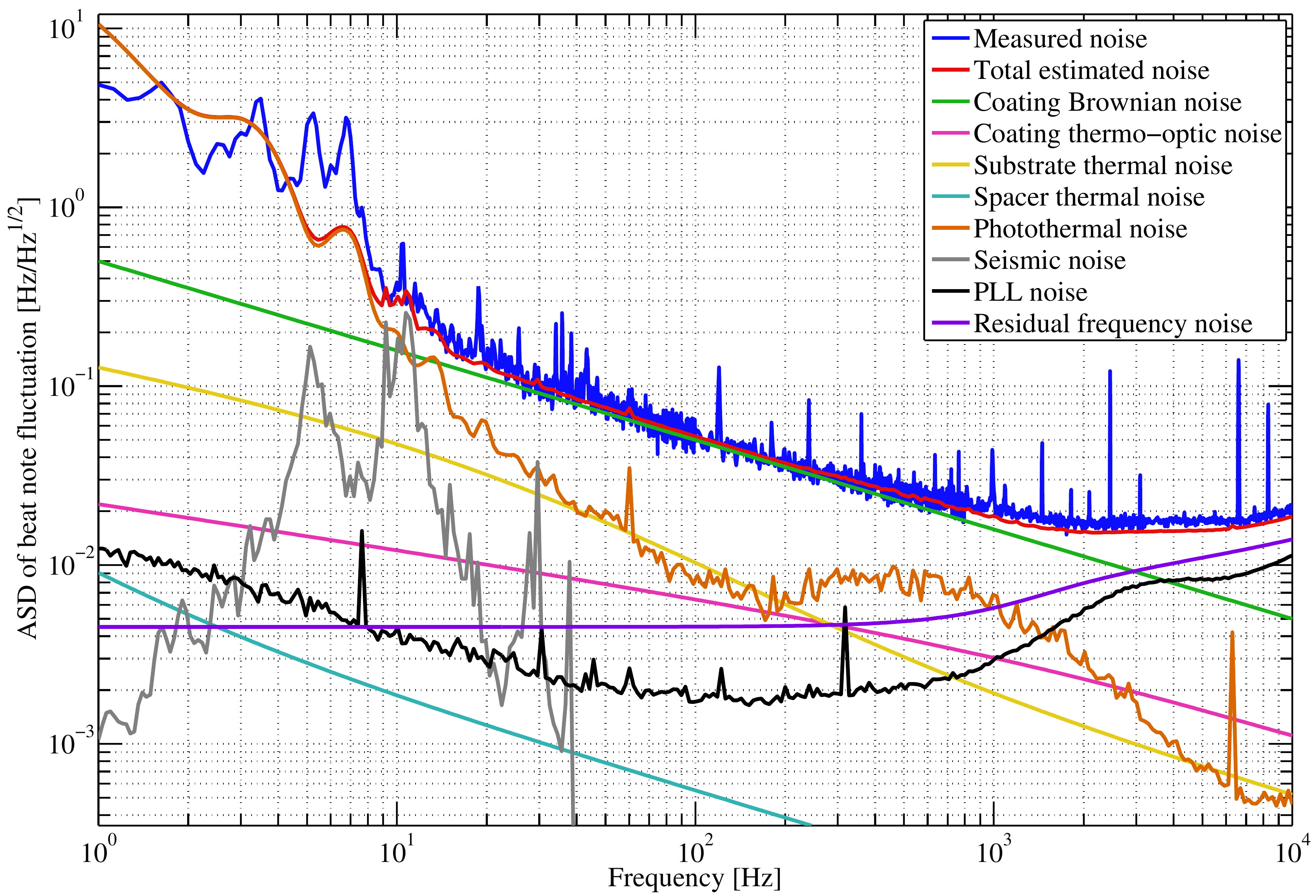}
    \caption{Amplitude spectral density $\sqrt{S_{\hat{\nu}}(f)}$ of beat note from two 3.7~cm cavities using the two-laser setup.
    In the band from 10~Hz to 1~kHz, the ASD has a $1/f^{1/2}$ slope with an amplitude consistent with coating Brownian noise.}
    \label{fig:nb_beat_short}
\end{figure*}
\begin{figure*}[tbp]
    \centering
    \includegraphics[width=0.85\textwidth]{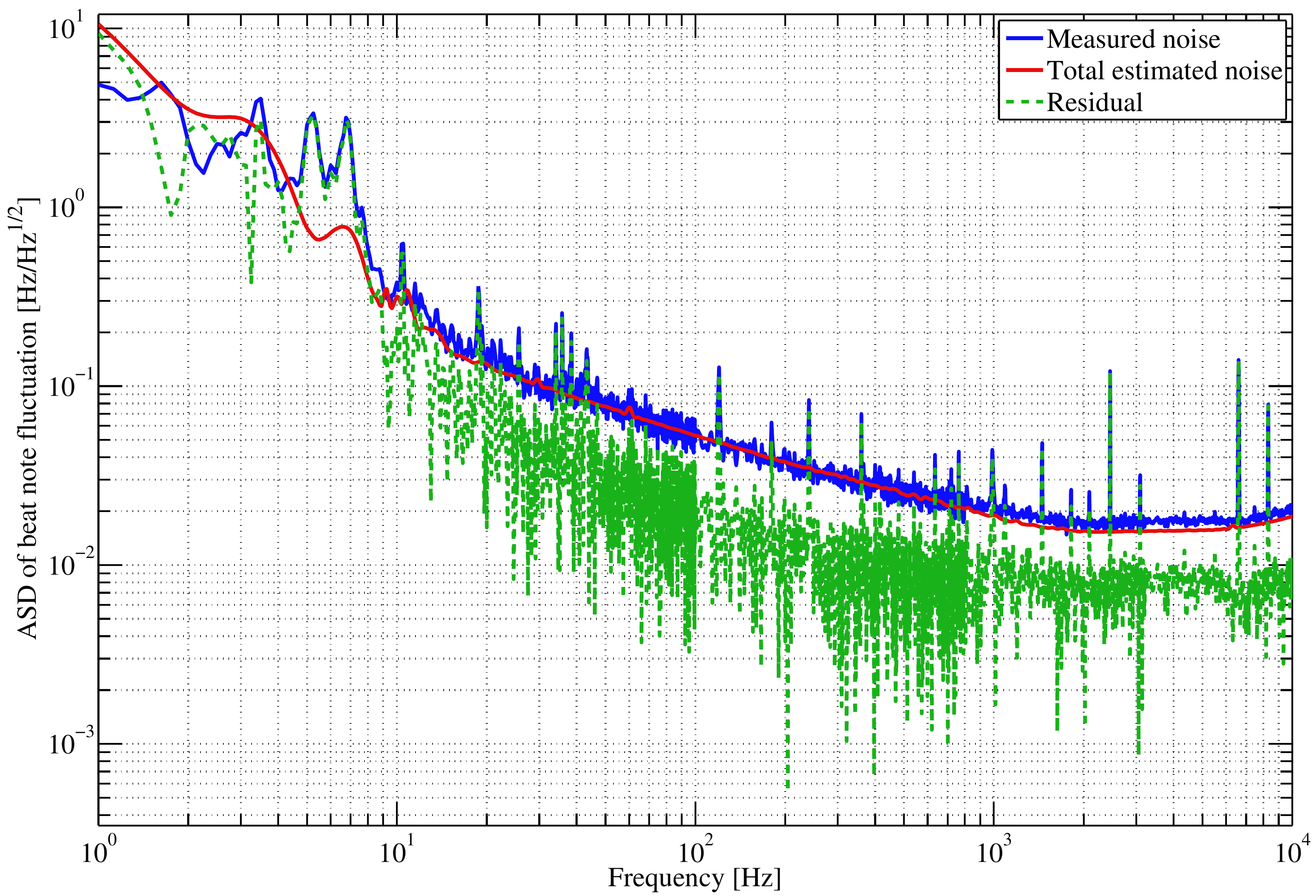}
    \caption{Measured and estimated amplitude spectral density $\sqrt{S_{\hat{\nu}}(f)}$ of beat note from 3.7\,cm cavities using the two-laser setup, along with the residual.
    With the exception of isolated peaks, above 10\,Hz the residual lies about a factor of two lower than the measured noise.}
    \label{fig:setup2_resid}
\end{figure*}

The beat note fluctuation of the 20.3\,cm cavities using the one-laser setup is shown in Figure~\ref{fig:nb_beat_long}.
In the band from 90\,--\,300\,Hz, the beat frequency noise has an amplitude spectral density $\sqrt{S_{\hat{\nu}}(f)}$ that is approximately $\bigl(8\times10^{-2} \text{ Hz}\bigr)/f^{1/2}$, although it is heavily contaminated by peaks.
Conversion into single-cavity length noise via $\sqrt{2}c/L\lambda$ gives $\sqrt{S_L(f)} = \bigl(4\times10^{-17} \text{ m}\bigr)/f^{1/2}$.

The beat note fluctuation of the 3.7\,cm cavities using the two-laser setup is shown in Figure~\ref{fig:nb_beat_short}, along with all the expected noise terms.
In the region from 10\,--\,1000\,Hz, the beat fluctuation has an ASD of approximately $\sqrt{S_{\hat{\nu}}(f)} = \bigl(0.5 \text{ Hz}\bigr)/f^{1/2}$, which is equvalent to $\sqrt{S_L(f)} = \bigl(5\times10^{-17} \text{ m}\bigr)/f^{1/2}$.

\subsubsection{Estimate of $\phi_\text{c}$}

We perform a fit for $\phi_\text{c}$ (defined in equation \ref{eq:Nakagawa_BR_coat}) in the region from 50\,--\,500\,Hz, where the measured ASD appears to be dominated by coating thermal noise.
We exclude bins near 60\,Hz and its harmonics.
We write the total estimated noise as $S_{\hat{\nu}}^{(\text{est})} = S_{\hat{\nu}}^{(\text{cBr})} + S_{\hat{\nu}}^{(\text{other})}$, where $S_{\hat{\nu}}^{(\text{cBr})}$ is determined from eq.~\ref{eq:Nakagawa_BR_coat}.
Then we perform a least squares fit of $S_{\hat{\nu}}^{(\text{meas})} - S_{\hat{\nu}}^{(\text{other})}$ to the functional form $A f^a$, for constant $A$ and $a$.
We find $A = (0.261\pm 0.015)\,\text{Hz}^2$ and $a = -1.004\pm0.011$.
Then from eq.~\ref{eq:Nakagawa_BR_coat}, we find $\phi_\text{c} = (4.43\pm0.25)\times10^{-4}$.

In Figure~\ref{fig:setup2_resid}, we plot the measured length noise, the total noise predicted from the noise budget, and the residual, found by performing the quadrature subtraction $\sqrt{S_{\hat{\nu}}^{(\text{resid})}} = \bigl|S_{\hat{\nu}}^{(\text{meas})} - S_{\hat{\nu}}^{(\text{est})}\bigr|^{1/2}$.

With the fitted loss angle, we calculate the coating thermal noise in the 20.3\,cm cavity, and plot it on the noise budget.
This is shown in Figure~\ref{fig:nb_beat_long}.
The measurement and the estimate total noise are in good agreement.
This is strong evidence that both measurements are dominated by coating thermal noise, since the amplitude of the PSD scales correctly with the spot size.
Additionally, our fitted loss angle $\phi_\text{c}$ is in good agreement with the results of Numata et~al.~\cite{Numata2003}, who found $\phi_\text{c} = 4\times10^{-4}$.
Finally, the shape of the beat note ASD for our two-cavity measurement is close to $f^{-1/2}$, as predicted by Eq.~\ref{eq:Nakagawa_BR_coat}.

\begin{figure*}
    \begin{subfigure}[tbp]{0.33\textwidth}
        \includegraphics[width=\textwidth]{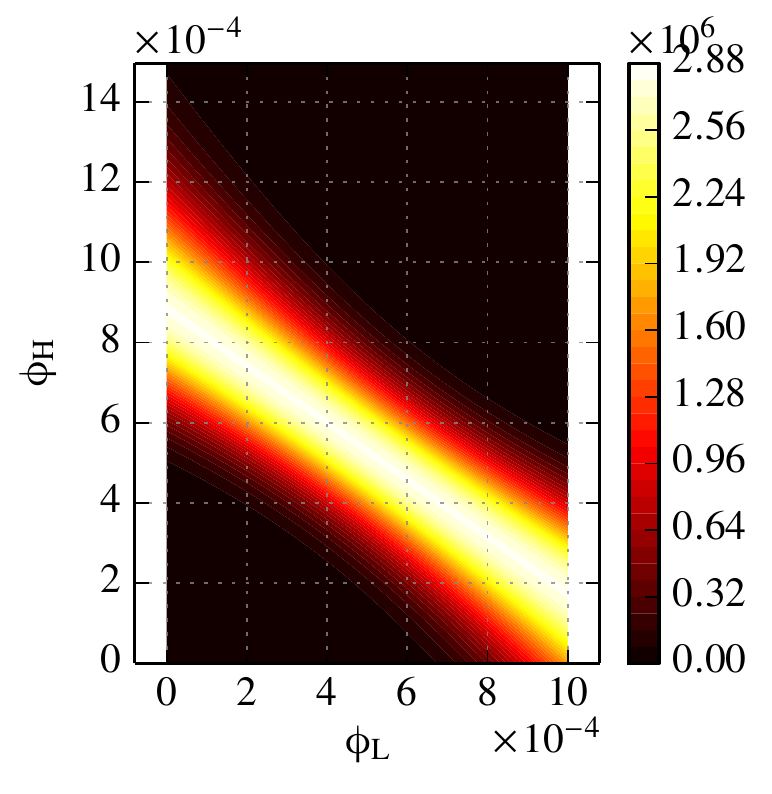}
        \caption{Prior PDF, after Harry et~al.~\cite{Harry2002}}
        \label{fig:prior}
    \end{subfigure}
    \begin{subfigure}[tbp]{0.33\textwidth}
        \includegraphics[width=\textwidth]{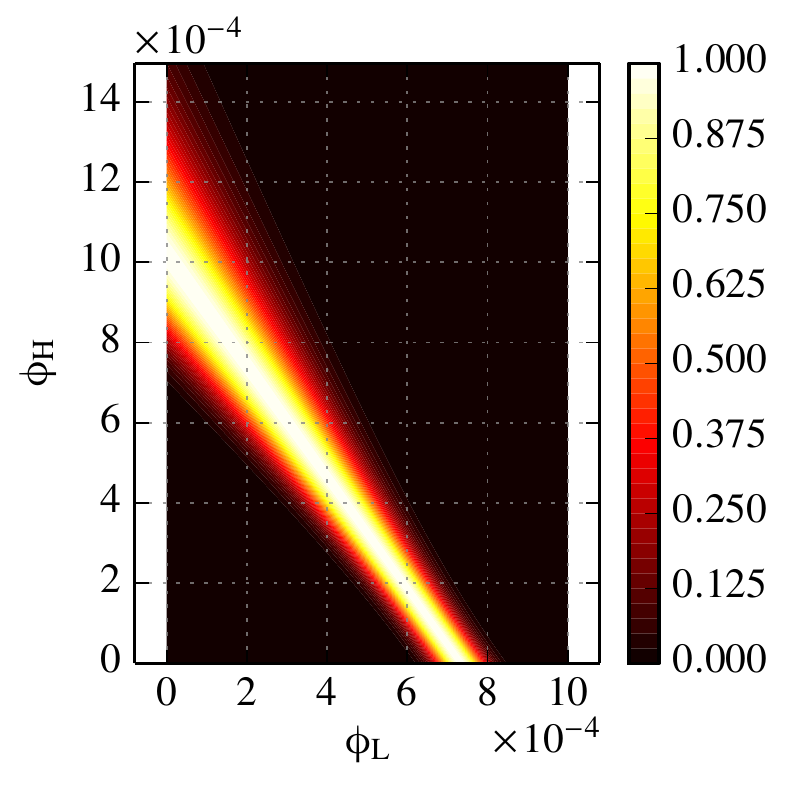}
        \caption{Likelihood}
        \label{fig:likelihood}
    \end{subfigure}
    \begin{subfigure}[tbp]{0.33\textwidth}
        \includegraphics[width=\textwidth]{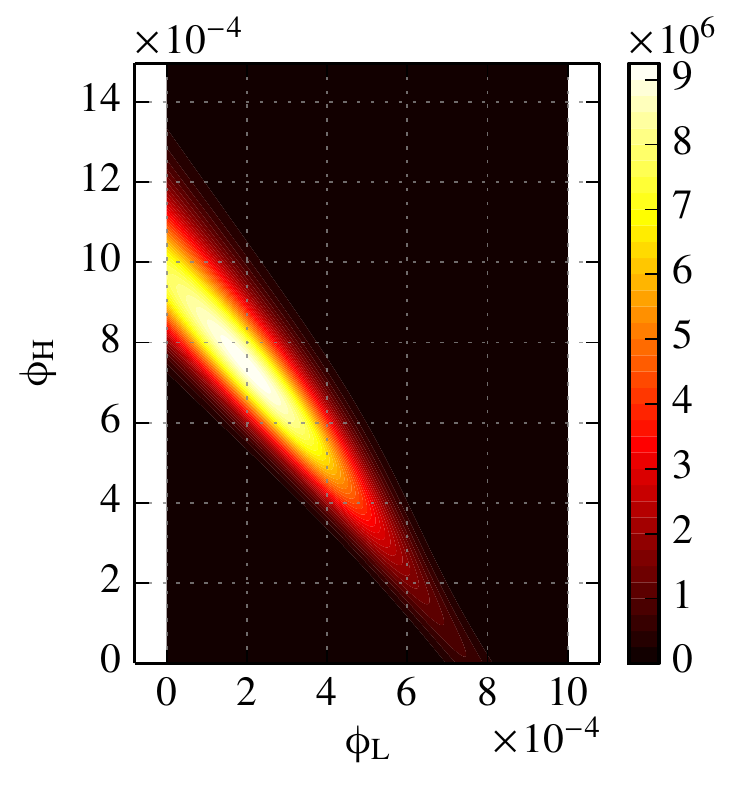}
        \caption{Posterior PDF}
        \label{fig:posterior}
    \end{subfigure}
    \caption{Prior PDF, likelihood, and posterior PDF used for Bayesian estimation of the loss angles of silica and tantala.}
    \label{fig:bayesian}
\end{figure*}

\begin{figure}
    \centering
    \includegraphics[width=0.45\textwidth]{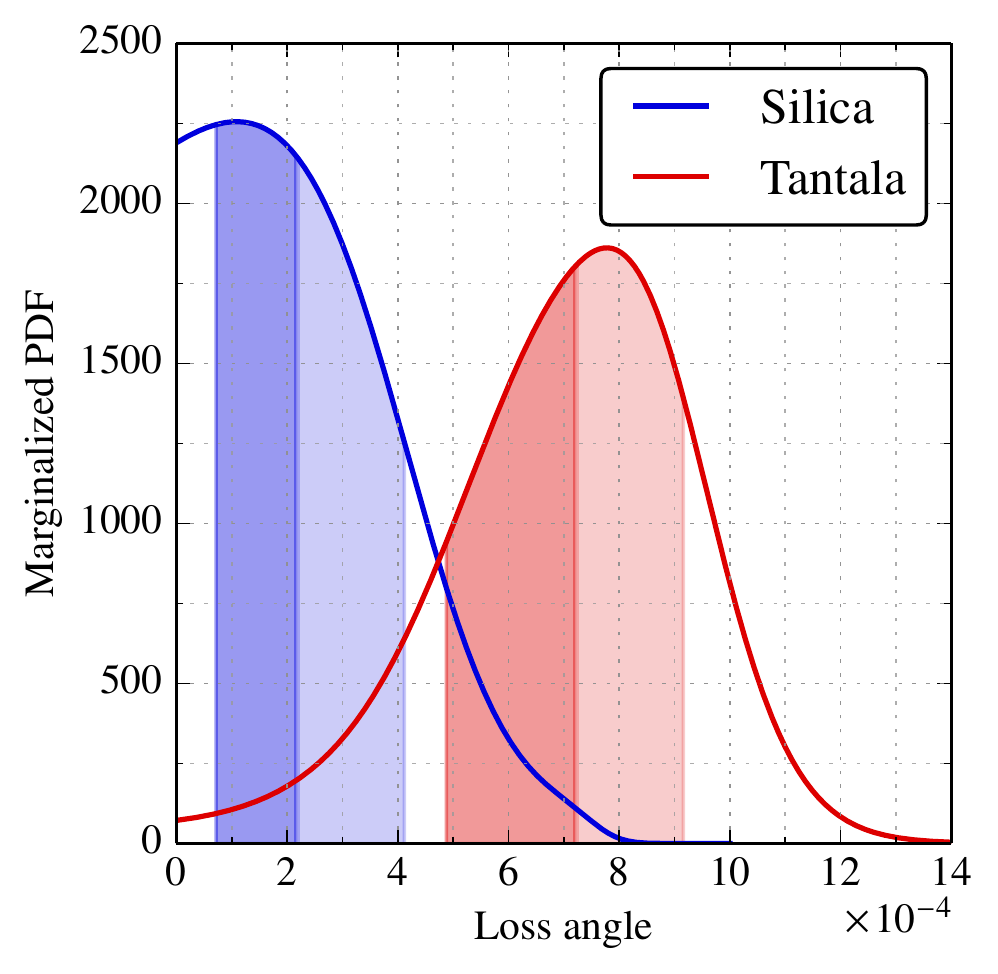}
    \caption{Marginalized posterior PDFs for $\phi_\text{L}$ (silica) and $\phi_\text{H}$ (tantala), with shaded regions demarcating the 16th, 50th, and 84th percentiles.}
    \label{fig:marginalized}
\end{figure}

\subsubsection{Estimate of $\phi_\text{L}$ and $\phi_\text{H}$}

Given $\phi_\text{c}$, a knowledge of the parameters of our coatings, and prior observations of coating loss angles, we can make a Bayesian estimate of $\phi_\text{L}$ and $\phi_\text{H}$.
To do this, we first write down a formula relating $\phi_\text{c}$, $\phi_\text{L}$ and $\phi_\text{H}$:
\begin{equation}
    \mathcal{M} \phi_\text{c} = \Xi_\text{L} N_\text{L} d_\text{L} \phi_\text{L} + \Xi_\text{H} N_\text{H} d_\text{H} \phi_\text{H}.
    \label{eq:numata_hong_relation}
\end{equation}
Here $\mathcal{M} = (1 + \sigma_\text{s})(1 - 2\sigma_\text{s}) d / E_\text{s}$, $N_\text{L} = 15$, $N_\text{H} = 14$, $d_\text{L} = \lambda/4n_\text{L}$, and $d_\text{H} = \lambda/4n_\text{H}$.
The coefficients $\Xi_\text{L}$ and $\Xi_\text{H}$ are found by combining Table~1, and Eqs.~94 and 96 from Hong et~al.~\cite{Hong2013}, assuming zero light penetration into the coating \footnote{For silica/tanala QWL coatings, most of the light penetrates only into the first few doublets}.
These coefficients depend only on the coating parameters.
Next we write down Bayes's theorem~\cite{vonToussaint2011}:
\begin{equation}
    p(\phi_\text{L}, \phi_\text{H} | \hat{\phi}_\text{c}) = \
    \frac{1}{Z}~\mathcal{L}(\phi_\text{L}, \phi_\text{H} | \hat{\phi}_\text{c})~p(\phi_\text{L}, \phi_\text{H}),
  \label{eq:bayes}
\end{equation}
where $Z$ is a normalization.
As a prior, we use data from the ringdown measurements in Harry et~al.~\cite{Harry2002}, since these measurements were performed on coatings from the same manufacturer as in our experiment, and were made during a similar time period.
Since Harry~et~al. performed a ringdown measurement, their quoted quantity $\phi_\parallel$ is distinct from $\phi_\text{c}$, and is related to the material loss angles $\phi_\text{L}$ and $\phi_\text{H}$ via
\begin{equation}
    (E_\text{L} d_\text{L} + E_\text{H} d_\text{H}) \phi_\parallel = E_\text{L} d_\text{L} \phi_\text{L} + E_\text{H} d_\text{H} \phi_\text{H}.
    \label{eq:phipara}
\end{equation}
We use $\hat{\phi}_\parallel \pm \sigma_{\hat{\phi}_\parallel} = (5.2 \pm 0.8)\times10^{-4}$ as the value measured by Harry et~al.~\footnote{Harry originally determined $\hat{\phi}_\parallel \pm \sigma_{\hat{\phi}_\parallel} = (1.0 \pm 0.3) \times 10^{-4}$ using a coating thickness that was 5 times the actual value.
Taking into account the correction given in Penn et~al.~\cite{Penn2003}, we reanalyze Harry's ringdown data to arrive at arrive at $\hat{\phi}_\parallel \pm \sigma_{\hat{\phi}_\parallel} = (5.2\pm0.8)\times10^{-4}$.}.

We then construct the prior
\begin{equation}
    p(\phi_\text{L}, \phi_\text{H}) = \frac{1}{Z_0}\exp\left[-\frac{1}{2} \frac{(\hat{\phi}_\parallel - \phi_\parallel)^2}{\sigma_{\hat{\phi}_\parallel}^2 + \sigma_{\phi_\parallel}^2}\right],
    \label{eq:prior}
\end{equation}
where $Z_0$ is a normalization, $\phi_\parallel$ is related to $\phi_\text{L}$ and $\phi_\text{H}$ via eq.~\ref{eq:phipara}, and $\sigma_{\phi_\parallel}$ is found by propagating forward the uncertainties on the material parameters as given in our Table~\ref{tab:cavity_params}.

As a likelihood we take
\begin{equation}
    \mathcal{L}(\phi_\text{L}, \phi_\text{H} | \hat{\phi}_\text{c}) = \exp\left[-\frac{1}{2} \frac{(\hat{\phi}_\text{c} - \phi_\text{c})^2}{\sigma_{\hat{\phi}_\text{c}}^2 + \sigma_{\phi_\text{c}}^2}\right]
    \label{eq:likelihood}
\end{equation}
with $\hat{\phi}_c$ given by our measurement, and $\phi_c$ given by 
Equation~\ref{eq:numata_hong_relation}.

The prior, the likelihood, and the resulting posterior are shown in Figures~\ref{fig:prior}--\ref{fig:posterior}.
In Figure~\ref{fig:marginalized}, we show the marginalized posterior PDFs for each loss angle.
For silica, we find the maximum \emph{a posteriori} (MAP) estimate for the loss angle $\phi_\text{L}$ is $1.1\times10^{-4}$, and the values for the 16th, 50th, and 84th percentiles are $0.7\times10^{-4}$, $2.2\times10^{-4}$, and $4.1\times10^{-4}$, respectively.
Likewise, for tantala, the MAP estimate for the loss angle $\phi_\text{H}$ is $7.8\times10^{-4}$, and the 16th, 50th, and 84th percentile values are $4.9\times10^{-4}$, $7.2\times10^{-4}$, and $9.2\times10^{-4}$, respectively.
The median (50th percentile) estimates for $\phi_\text{L}$ and $\phi_\text{H}$ are in agreement with the values that result from treating eqs.~\ref{eq:numata_hong_relation} and \ref{eq:phipara} as a system of two equations in two unknowns and solving for $\phi_\text{L}$ and $\phi_\text{H}$ (and propagating uncertainties accordingly); the results are $(2.0\pm2.2)\times10^{-4}$ and $(7.4\pm2.7)\times10^{-4}$, respectively.

\section{Conclusions}
\label{sec:conclusion}

\begin{figure}[tbph]
    \begin{subfigure}[tbph]{0.5\textwidth}
        \includegraphics[width=\textwidth]{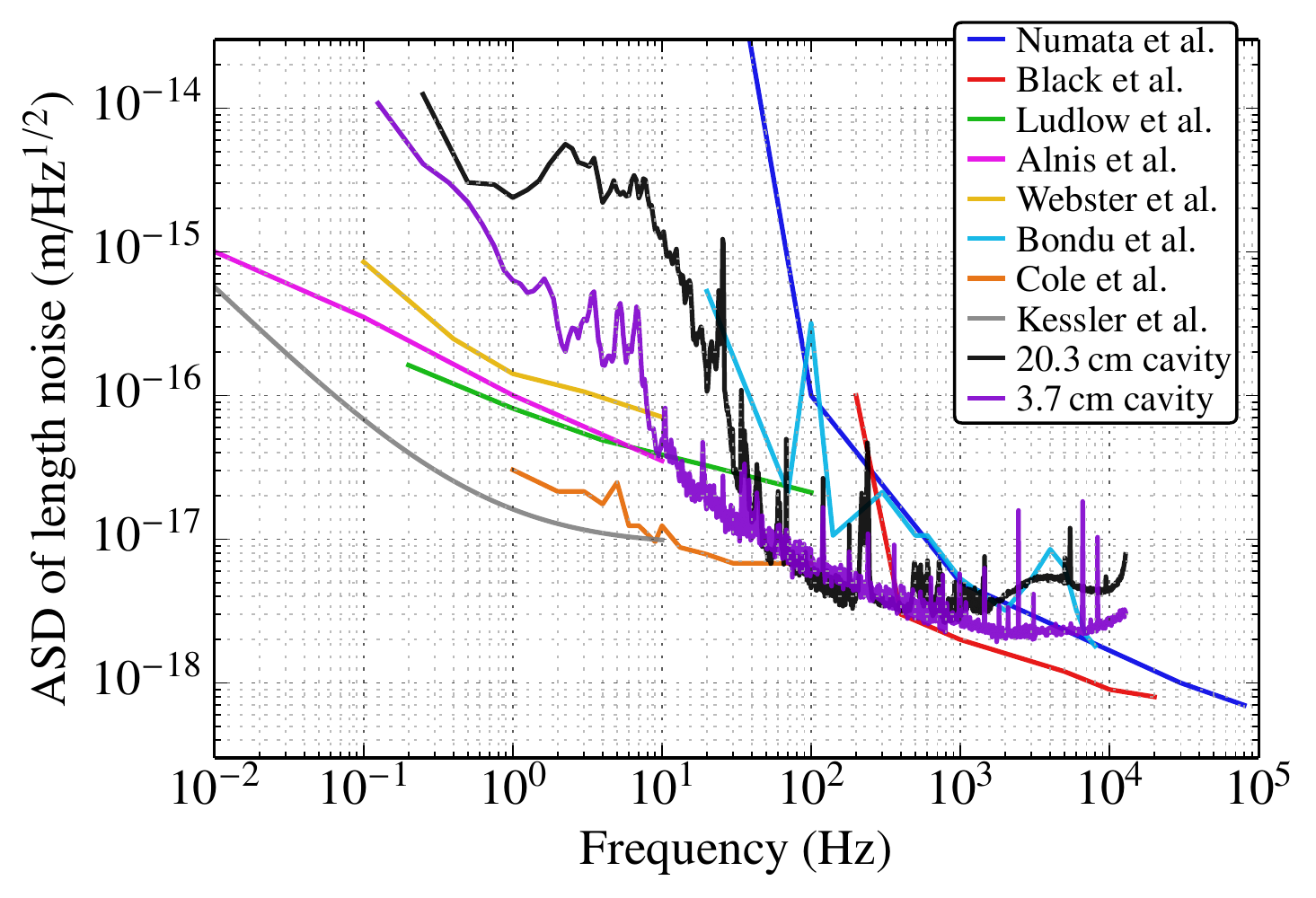}
    \end{subfigure}
    \begin{subfigure}[tbph]{0.5\textwidth}
        \includegraphics[width=\textwidth]{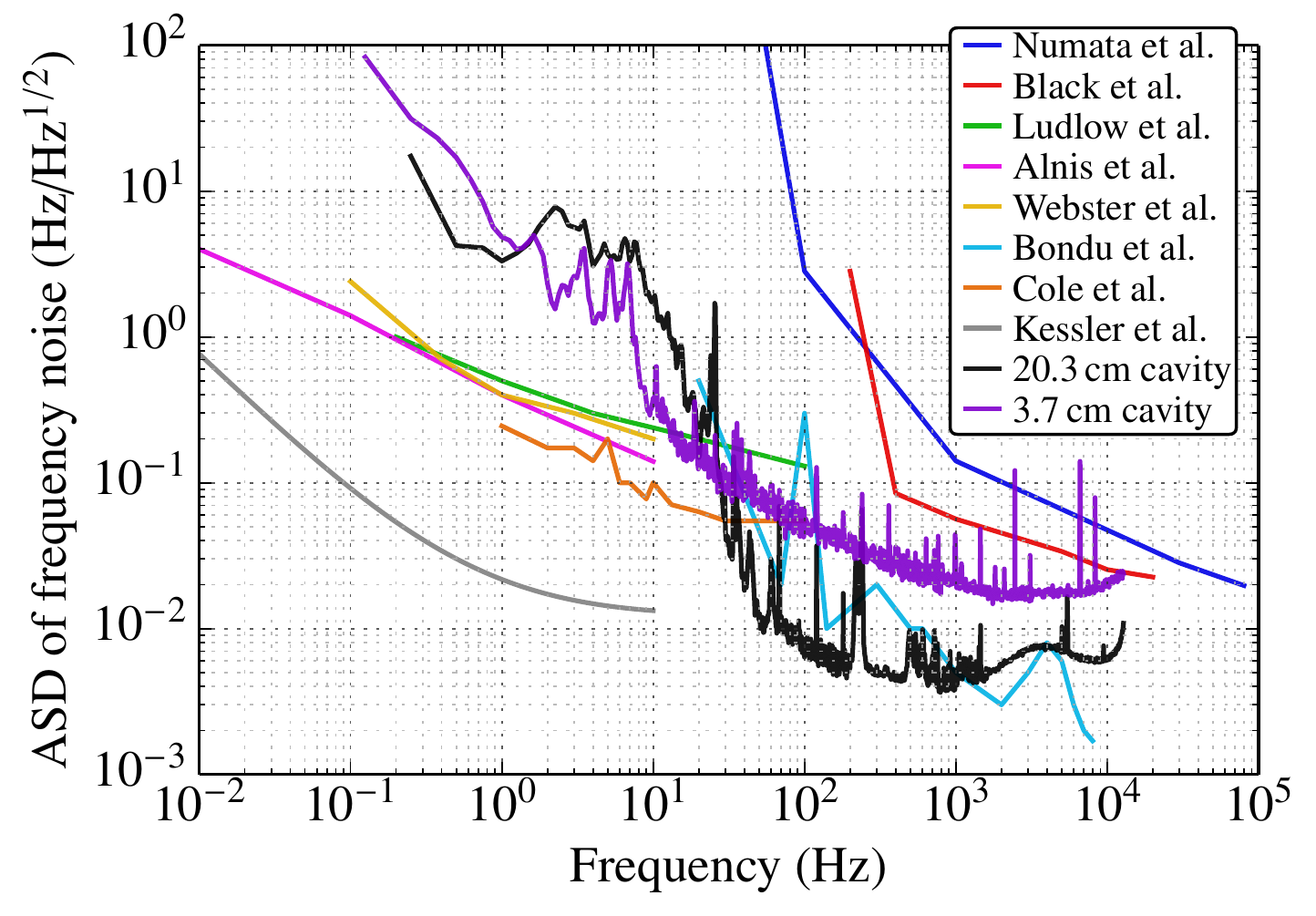}
    \end{subfigure}
    \caption{Comparison of the measurements in this work with measurements from other reference cavity experiments, both in terms of length noise (top) and frequency noise (bottom).
    The traces show measurements from Numata~et~al.~\cite{Numata2003}, Black~et~al.~\cite{Black2004_1}, Ludlow~et~al.~\cite{Ludlow2007}, Alnis~et~al.~\cite{Alnis2008}, Webster~et~al.~\cite{Webster2007b}, Cole~et~al.~\cite{Cole2013}, Kessler~et~al.~\cite{Kessler2012b}, and the two measurements presented in this work.}
    \label{fig:comparisonplots}
\end{figure}
    


In summary, we have demonstrated a high-sensitivity system for measuring thermal noise from high-reflectivity silica/tantala coatings.
The cavity spacers, the isolation system, the laser stabilization system, and the beat readout have been designed to push down known technical and environmental noise sources to below the expected thermal noise level.
Indeed, in the band from 10\,Hz to 1\,kHz, the measured beat spectrum produces length fluctuation consistent with Brownian noise from the mirror coatings, with a loss angle of $\phi_\text{c} = 4\times10^{-4}$.

The estimated loss angles $\phi_\text{L}$ and $\phi_\text{H}$ are high compared to other measurements given in the literature.
For example, ringdown measurements by Penn et~al.~\cite{Penn2003}, Crooks et~al.~\cite{Crooks2004, Crooks2006}, and Li et~al.~\cite{Li2014} found $\phi_\text{L} < 1\times10^{-4}$ and $\phi_\text{H} \sim 4\times10^{-4}$.
It is possible that these newer coatings were manufactured with better fabrication techniques, as manufacturers became more aware of coating thermal noise.

In Figure~\ref{fig:comparisonplots}, we plot our measurements, along with measurements from other reference cavities, in both displacement and frequency noise.
While other measurements have focused on attaining thermally limited noise performance at low frequencies (100\,Hz down to less than 10\,mHz) or high frequencies (100\,Hz up to 100\,kHz), the measurements presented in this paper are consistent with the thermal noise limit in an intermediate frequency band, from 10\,Hz to 1\,kHz, which is of direct interest to the current and future generations of gravitational wave detectors.

Brownian noise in optical coatings is a significant limit in precision optical measurements because of the high mechanical loss angle in the amorphous coatings.
Recent efforts have now begun to focus on other coating materials, such as monocrystalline Al$_x$Ga$_{1-x}$As heterostructure (AlGaAs). 
Measurements by Cole et~al.~\cite{Cole2013} on AlGaAs coatings with quarter-wavelength structures indicate the potential for a smaller thermal noise by almost an order of magnitude compared to that of silica/tantala coatings.
The current sensitivity of the testbed mentioned in this paper will need several improvements to be able to measure thermal noise in AlGaAs coating with better SNR. 
The RIN suppression servo will need to be upgraded to reduce RIN-induced photothermal noise from DC to 30\,Hz. 
To overcome the readout noise from the PLL at frequencies above 1\,kHz, another means of frequency noise detection (for example, a homodyne detection system~\cite{Eichholz2013, hartman2014measurement}) may need to be considered. 

Since AlGaAs coatings may be used in third-generation gravitational-wave detectors~\cite{Rana:RMP2014}, it is important to characterize all fundamental noises associated with the coatings to thoroughly estimate the detector's sensitivity. 
There are still several issues regarding thermal noise calculation in AlGaAs coatings to be explored. 
First, the current theoretical calculations~\cite{Hong2013} of coating Brownian noise may have to be revised to include tensorial components in the elasticity equations; the current calculations assume that coating properties are isotropic in the amorphous thin films.
Second, thermo-optic noise in AlGaAs coatings is predicted to be significant due to its high thermo-refactive coefficient. 
Since GaAs and AlAs have a high thermal conductivity compared to silica/tantala coatings, the assumptions used by Evans et~al.~\cite{Evans2008} to compute thermo-optic noise will no longer be accurate: corrections for a small spot size and low frequencies will be required.
In addition, it seems possible to minimize thermo-optic noise by adjusting the crystalline coating structure~\cite{AlGaAs:TO}, so that the limiting noise floor of the coating can be further reduced.

\begin{acknowledgments}
The authors wish to thank Rich Abbott, Daniel Sigg, David Yeaton-Massey, Larry Price, Peter King, Matt Abernathy, Megan Daily, Raphael Cervantes, Sarah Terry, and Nicolas Smith-Lefebvre for useful discussions and technical help.
We thank Borja Sorazu for careful reading of the manuscript.

LIGO was constructed by the California Institute of Technology and Massachusetts Institute of Technology with funding from the National Science Foundation and operates under cooperative agreement PHY-0757058. 
This paper has LIGO Document Number LIGO-P1400072.
\end{acknowledgments}

\nocite{Bondu1996, Kessler2012b} 
\bibliography{bib/ctnbib.bib}

\begin{thebibliography}{50}%
\makeatletter
\providecommand \@ifxundefined [1]{%
 \@ifx{#1\undefined}
}%
\providecommand \@ifnum [1]{%
 \ifnum #1\expandafter \@firstoftwo
 \else \expandafter \@secondoftwo
 \fi
}%
\providecommand \@ifx [1]{%
 \ifx #1\expandafter \@firstoftwo
 \else \expandafter \@secondoftwo
 \fi
}%
\providecommand \natexlab [1]{#1}%
\providecommand \enquote  [1]{``#1''}%
\providecommand \bibnamefont  [1]{#1}%
\providecommand \bibfnamefont [1]{#1}%
\providecommand \citenamefont [1]{#1}%
\providecommand \href@noop [0]{\@secondoftwo}%
\providecommand \href [0]{\begingroup \@sanitize@url \@href}%
\providecommand \@href[1]{\@@startlink{#1}\@@href}%
\providecommand \@@href[1]{\endgroup#1\@@endlink}%
\providecommand \@sanitize@url [0]{\catcode `\\12\catcode `\$12\catcode
  `\&12\catcode `\#12\catcode `\^12\catcode `\_12\catcode `\%12\relax}%
\providecommand \@@startlink[1]{}%
\providecommand \@@endlink[0]{}%
\providecommand \url  [0]{\begingroup\@sanitize@url \@url }%
\providecommand \@url [1]{\endgroup\@href {#1}{\urlprefix }}%
\providecommand \urlprefix  [0]{URL }%
\providecommand \Eprint [0]{\href }%
\providecommand \doibase [0]{http://dx.doi.org/}%
\providecommand \selectlanguage [0]{\@gobble}%
\providecommand \bibinfo  [0]{\@secondoftwo}%
\providecommand \bibfield  [0]{\@secondoftwo}%
\providecommand \translation [1]{[#1]}%
\providecommand \BibitemOpen [0]{}%
\providecommand \bibitemStop [0]{}%
\providecommand \bibitemNoStop [0]{.\EOS\space}%
\providecommand \EOS [0]{\spacefactor3000\relax}%
\providecommand \BibitemShut  [1]{\csname bibitem#1\endcsname}%
\let\auto@bib@innerbib\@empty
\bibitem [{\citenamefont {Harry}\ and\ \citenamefont {the LIGO
  Scientific~Collaboration}(2010)}]{Harry2010}%
  \BibitemOpen
  \bibfield  {author} {\bibinfo {author} {\bibfnamefont {G.~M.}\ \bibnamefont
  {Harry}}\ and\ \bibinfo {author} {\bibnamefont {the LIGO
  Scientific~Collaboration}},\ }\href
  {http://stacks.iop.org/0264-9381/27/i=8/a=084006} {\bibfield  {journal}
  {\bibinfo  {journal} {Classical and Quantum Gravity}\ }\textbf {\bibinfo
  {volume} {27}},\ \bibinfo {pages} {084006} (\bibinfo {year}
  {2010})}\BibitemShut {NoStop}%
\bibitem [{\citenamefont {Nakagawa}\ \emph {et~al.}(2002)\citenamefont
  {Nakagawa}, \citenamefont {Gretarsson}, \citenamefont {Gustafson},\ and\
  \citenamefont {Fejer}}]{Nakagawa2002}%
  \BibitemOpen
  \bibfield  {author} {\bibinfo {author} {\bibfnamefont {N.}~\bibnamefont
  {Nakagawa}}, \bibinfo {author} {\bibfnamefont {A.~M.}\ \bibnamefont
  {Gretarsson}}, \bibinfo {author} {\bibfnamefont {E.~K.}\ \bibnamefont
  {Gustafson}}, \ and\ \bibinfo {author} {\bibfnamefont {M.~M.}\ \bibnamefont
  {Fejer}},\ }\href {\doibase 10.1103/PhysRevD.65.102001} {\bibfield  {journal}
  {\bibinfo  {journal} {Phys. Rev. D}\ }\textbf {\bibinfo {volume} {65}},\
  \bibinfo {pages} {102001} (\bibinfo {year} {2002})}\BibitemShut {NoStop}%
\bibitem [{\citenamefont {Harry}\ \emph {et~al.}(2002)\citenamefont {Harry},
  \citenamefont {Gretarsson}, \citenamefont {Saulson}, \citenamefont
  {Kittelberger}, \citenamefont {Penn}, \citenamefont {Startin}, \citenamefont
  {Rowan}, \citenamefont {Fejer}, \citenamefont {Crooks}, \citenamefont
  {Cagnoli}, \citenamefont {Hough},\ and\ \citenamefont
  {Nakagawa}}]{Harry2002}%
  \BibitemOpen
  \bibfield  {author} {\bibinfo {author} {\bibfnamefont {G.~M.}\ \bibnamefont
  {Harry}}, \bibinfo {author} {\bibfnamefont {A.~M.}\ \bibnamefont
  {Gretarsson}}, \bibinfo {author} {\bibfnamefont {P.~R.}\ \bibnamefont
  {Saulson}}, \bibinfo {author} {\bibfnamefont {S.~E.}\ \bibnamefont
  {Kittelberger}}, \bibinfo {author} {\bibfnamefont {S.~D.}\ \bibnamefont
  {Penn}}, \bibinfo {author} {\bibfnamefont {W.~J.}\ \bibnamefont {Startin}},
  \bibinfo {author} {\bibfnamefont {S.}~\bibnamefont {Rowan}}, \bibinfo
  {author} {\bibfnamefont {M.~M.}\ \bibnamefont {Fejer}}, \bibinfo {author}
  {\bibfnamefont {D.~R.~M.}\ \bibnamefont {Crooks}}, \bibinfo {author}
  {\bibfnamefont {G.}~\bibnamefont {Cagnoli}}, \bibinfo {author} {\bibfnamefont
  {J.}~\bibnamefont {Hough}}, \ and\ \bibinfo {author} {\bibfnamefont
  {N.}~\bibnamefont {Nakagawa}},\ }\href
  {http://stacks.iop.org/0264-9381/19/i=5/a=305} {\bibfield  {journal}
  {\bibinfo  {journal} {Classical and Quantum Gravity}\ }\textbf {\bibinfo
  {volume} {19}},\ \bibinfo {pages} {897} (\bibinfo {year} {2002})}\BibitemShut
  {NoStop}%
\bibitem [{\citenamefont {Somiya}\ and\ \citenamefont
  {Yamamoto}(2009)}]{Somiya2009}%
  \BibitemOpen
  \bibfield  {author} {\bibinfo {author} {\bibfnamefont {K.}~\bibnamefont
  {Somiya}}\ and\ \bibinfo {author} {\bibfnamefont {K.}~\bibnamefont
  {Yamamoto}},\ }\href {\doibase 10.1103/PhysRevD.79.102004} {\bibfield
  {journal} {\bibinfo  {journal} {Phys. Rev. D}\ }\textbf {\bibinfo {volume}
  {79}},\ \bibinfo {pages} {102004} (\bibinfo {year} {2009})}\BibitemShut
  {NoStop}%
\bibitem [{\citenamefont {Hong}\ \emph {et~al.}(2013)\citenamefont {Hong},
  \citenamefont {Yang}, \citenamefont {Gustafson}, \citenamefont {Adhikari},\
  and\ \citenamefont {Chen}}]{Hong2013}%
  \BibitemOpen
  \bibfield  {author} {\bibinfo {author} {\bibfnamefont {T.}~\bibnamefont
  {Hong}}, \bibinfo {author} {\bibfnamefont {H.}~\bibnamefont {Yang}}, \bibinfo
  {author} {\bibfnamefont {E.~K.}\ \bibnamefont {Gustafson}}, \bibinfo {author}
  {\bibfnamefont {R.~X.}\ \bibnamefont {Adhikari}}, \ and\ \bibinfo {author}
  {\bibfnamefont {Y.}~\bibnamefont {Chen}},\ }\href {\doibase
  10.1103/PhysRevD.87.082001} {\bibfield  {journal} {\bibinfo  {journal} {Phys.
  Rev. D}\ }\textbf {\bibinfo {volume} {87}},\ \bibinfo {pages} {082001}
  (\bibinfo {year} {2013})}\BibitemShut {NoStop}%
\bibitem [{\citenamefont {Numata}\ \emph {et~al.}(2003)\citenamefont {Numata},
  \citenamefont {Ando}, \citenamefont {Yamamoto}, \citenamefont {Otsuka},\ and\
  \citenamefont {Tsubono}}]{Numata2003}%
  \BibitemOpen
  \bibfield  {author} {\bibinfo {author} {\bibfnamefont {K.}~\bibnamefont
  {Numata}}, \bibinfo {author} {\bibfnamefont {M.}~\bibnamefont {Ando}},
  \bibinfo {author} {\bibfnamefont {K.}~\bibnamefont {Yamamoto}}, \bibinfo
  {author} {\bibfnamefont {S.}~\bibnamefont {Otsuka}}, \ and\ \bibinfo {author}
  {\bibfnamefont {K.}~\bibnamefont {Tsubono}},\ }\href {\doibase
  10.1103/PhysRevLett.91.260602} {\bibfield  {journal} {\bibinfo  {journal}
  {Phys. Rev. Lett.}\ }\textbf {\bibinfo {volume} {91}},\ \bibinfo {pages}
  {260602} (\bibinfo {year} {2003})}\BibitemShut {NoStop}%
\bibitem [{\citenamefont {Black}\ \emph {et~al.}(2004)\citenamefont {Black},
  \citenamefont {Villar}, \citenamefont {Barbary}, \citenamefont {Bushmaker},
  \citenamefont {Heefner}, \citenamefont {Kawamura}, \citenamefont {Kawazoe},
  \citenamefont {Matone}, \citenamefont {Meidt}, \citenamefont {Rao},
  \citenamefont {Schulz}, \citenamefont {Zhang},\ and\ \citenamefont
  {Libbrecht}}]{Black2004_1}%
  \BibitemOpen
  \bibfield  {author} {\bibinfo {author} {\bibfnamefont {E.~D.}\ \bibnamefont
  {Black}}, \bibinfo {author} {\bibfnamefont {A.}~\bibnamefont {Villar}},
  \bibinfo {author} {\bibfnamefont {K.}~\bibnamefont {Barbary}}, \bibinfo
  {author} {\bibfnamefont {A.}~\bibnamefont {Bushmaker}}, \bibinfo {author}
  {\bibfnamefont {J.}~\bibnamefont {Heefner}}, \bibinfo {author} {\bibfnamefont
  {S.}~\bibnamefont {Kawamura}}, \bibinfo {author} {\bibfnamefont
  {F.}~\bibnamefont {Kawazoe}}, \bibinfo {author} {\bibfnamefont
  {L.}~\bibnamefont {Matone}}, \bibinfo {author} {\bibfnamefont
  {S.}~\bibnamefont {Meidt}}, \bibinfo {author} {\bibfnamefont {S.~R.}\
  \bibnamefont {Rao}}, \bibinfo {author} {\bibfnamefont {K.}~\bibnamefont
  {Schulz}}, \bibinfo {author} {\bibfnamefont {M.}~\bibnamefont {Zhang}}, \
  and\ \bibinfo {author} {\bibfnamefont {K.~G.}\ \bibnamefont {Libbrecht}},\
  }\href {\doibase 10.1016/j.physleta.2004.05.052} {\bibfield  {journal}
  {\bibinfo  {journal} {Physics Letters A}\ }\textbf {\bibinfo {volume}
  {328}},\ \bibinfo {pages} {1 } (\bibinfo {year} {2004})}\BibitemShut
  {NoStop}%
\bibitem [{\citenamefont {Ludlow}\ \emph {et~al.}(2007)\citenamefont {Ludlow},
  \citenamefont {Huang}, \citenamefont {Notcutt}, \citenamefont
  {Zanon-Willette}, \citenamefont {Foreman}, \citenamefont {Boyd},
  \citenamefont {Blatt},\ and\ \citenamefont {Ye}}]{Ludlow2007}%
  \BibitemOpen
  \bibfield  {author} {\bibinfo {author} {\bibfnamefont {A.~D.}\ \bibnamefont
  {Ludlow}}, \bibinfo {author} {\bibfnamefont {X.}~\bibnamefont {Huang}},
  \bibinfo {author} {\bibfnamefont {M.}~\bibnamefont {Notcutt}}, \bibinfo
  {author} {\bibfnamefont {T.}~\bibnamefont {Zanon-Willette}}, \bibinfo
  {author} {\bibfnamefont {S.~M.}\ \bibnamefont {Foreman}}, \bibinfo {author}
  {\bibfnamefont {M.~M.}\ \bibnamefont {Boyd}}, \bibinfo {author}
  {\bibfnamefont {S.}~\bibnamefont {Blatt}}, \ and\ \bibinfo {author}
  {\bibfnamefont {J.}~\bibnamefont {Ye}},\ }\href {\doibase
  10.1364/OL.32.000641} {\bibfield  {journal} {\bibinfo  {journal} {Opt.
  Lett.}\ }\textbf {\bibinfo {volume} {32}},\ \bibinfo {pages} {641} (\bibinfo
  {year} {2007})}\BibitemShut {NoStop}%
\bibitem [{\citenamefont {Alnis}\ \emph {et~al.}(2008)\citenamefont {Alnis},
  \citenamefont {Matveev}, \citenamefont {Kolachevsky}, \citenamefont {Udem},\
  and\ \citenamefont {H\"ansch}}]{Alnis2008}%
  \BibitemOpen
  \bibfield  {author} {\bibinfo {author} {\bibfnamefont {J.}~\bibnamefont
  {Alnis}}, \bibinfo {author} {\bibfnamefont {A.}~\bibnamefont {Matveev}},
  \bibinfo {author} {\bibfnamefont {N.}~\bibnamefont {Kolachevsky}}, \bibinfo
  {author} {\bibfnamefont {T.}~\bibnamefont {Udem}}, \ and\ \bibinfo {author}
  {\bibfnamefont {T.~W.}\ \bibnamefont {H\"ansch}},\ }\href {\doibase
  10.1103/PhysRevA.77.053809} {\bibfield  {journal} {\bibinfo  {journal} {Phys.
  Rev. A}\ }\textbf {\bibinfo {volume} {77}},\ \bibinfo {pages} {053809}
  (\bibinfo {year} {2008})}\BibitemShut {NoStop}%
\bibitem [{\citenamefont {Webster}\ \emph {et~al.}(2008)\citenamefont
  {Webster}, \citenamefont {Oxborrow}, \citenamefont {Pugla}, \citenamefont
  {Millo},\ and\ \citenamefont {Gill}}]{Webster2007b}%
  \BibitemOpen
  \bibfield  {author} {\bibinfo {author} {\bibfnamefont {S.~A.}\ \bibnamefont
  {Webster}}, \bibinfo {author} {\bibfnamefont {M.}~\bibnamefont {Oxborrow}},
  \bibinfo {author} {\bibfnamefont {S.}~\bibnamefont {Pugla}}, \bibinfo
  {author} {\bibfnamefont {J.}~\bibnamefont {Millo}}, \ and\ \bibinfo {author}
  {\bibfnamefont {P.}~\bibnamefont {Gill}},\ }\href {\doibase
  10.1103/PhysRevA.77.033847} {\bibfield  {journal} {\bibinfo  {journal} {Phys.
  Rev. A}\ }\textbf {\bibinfo {volume} {77}},\ \bibinfo {pages} {033847}
  (\bibinfo {year} {2008})}\BibitemShut {NoStop}%
\bibitem [{\citenamefont {Callen}\ and\ \citenamefont
  {Greene}(1952)}]{Callen1952}%
  \BibitemOpen
  \bibfield  {author} {\bibinfo {author} {\bibfnamefont {H.~B.}\ \bibnamefont
  {Callen}}\ and\ \bibinfo {author} {\bibfnamefont {R.~F.}\ \bibnamefont
  {Greene}},\ }\href@noop {} {\bibfield  {journal} {\bibinfo  {journal}
  {Physical Review}\ }\textbf {\bibinfo {volume} {86}} (\bibinfo {year}
  {1952})}\BibitemShut {NoStop}%
\bibitem [{\citenamefont {Saulson}(1990)}]{Saulson1990}%
  \BibitemOpen
  \bibfield  {author} {\bibinfo {author} {\bibfnamefont {P.~R.}\ \bibnamefont
  {Saulson}},\ }\href@noop {} {\bibfield  {journal} {\bibinfo  {journal}
  {Physical Review D}\ }\textbf {\bibinfo {volume} {42}} (\bibinfo {year}
  {1990})}\BibitemShut {NoStop}%
\bibitem [{\citenamefont {Levin}(1998)}]{Levin1998}%
  \BibitemOpen
  \bibfield  {author} {\bibinfo {author} {\bibfnamefont {{\relax
  Yu}.}~\bibnamefont {Levin}},\ }\href {\doibase 10.1103/PhysRevD.57.659}
  {\bibfield  {journal} {\bibinfo  {journal} {Phys. Rev. D}\ }\textbf {\bibinfo
  {volume} {57}},\ \bibinfo {pages} {659} (\bibinfo {year} {1998})}\BibitemShut
  {NoStop}%
\bibitem [{\citenamefont {Gonzales}\ and\ \citenamefont
  {Saulson}(1994)}]{Gonzalez1994}%
  \BibitemOpen
  \bibfield  {author} {\bibinfo {author} {\bibfnamefont {G.~I.}\ \bibnamefont
  {Gonzales}}\ and\ \bibinfo {author} {\bibfnamefont {P.~R.}\ \bibnamefont
  {Saulson}},\ }\href@noop {} {\bibfield  {journal} {\bibinfo  {journal} {J.
  Acoust. Soc. Am.}\ }\textbf {\bibinfo {volume} {96}} (\bibinfo {year}
  {1994})}\BibitemShut {NoStop}%
\bibitem [{\citenamefont {Zener}(1938)}]{Zener1938}%
  \BibitemOpen
  \bibfield  {author} {\bibinfo {author} {\bibfnamefont {C.}~\bibnamefont
  {Zener}},\ }\href {\doibase 10.1103/PhysRev.53.90} {\bibfield  {journal}
  {\bibinfo  {journal} {Phys. Rev.}\ }\textbf {\bibinfo {volume} {53}},\
  \bibinfo {pages} {90} (\bibinfo {year} {1938})}\BibitemShut {NoStop}%
\bibitem [{\citenamefont {Liu}\ and\ \citenamefont {Thorne}(2000)}]{Liu2000}%
  \BibitemOpen
  \bibfield  {author} {\bibinfo {author} {\bibfnamefont {Y.~T.}\ \bibnamefont
  {Liu}}\ and\ \bibinfo {author} {\bibfnamefont {K.~S.}\ \bibnamefont
  {Thorne}},\ }\href {\doibase 10.1103/PhysRevD.62.122002} {\bibfield
  {journal} {\bibinfo  {journal} {Phys. Rev. D}\ }\textbf {\bibinfo {volume}
  {62}},\ \bibinfo {pages} {122002} (\bibinfo {year} {2000})}\BibitemShut
  {NoStop}%
\bibitem [{\citenamefont {Levin}(2008)}]{Levin2008}%
  \BibitemOpen
  \bibfield  {author} {\bibinfo {author} {\bibfnamefont {Y.}~\bibnamefont
  {Levin}},\ }\href {\doibase http://dx.doi.org/10.1016/j.physleta.2007.11.007}
  {\bibfield  {journal} {\bibinfo  {journal} {Physics Letters A}\ }\textbf
  {\bibinfo {volume} {372}},\ \bibinfo {pages} {1941 } (\bibinfo {year}
  {2008})}\BibitemShut {NoStop}%
\bibitem [{\citenamefont {Evans}\ \emph {et~al.}(2008)\citenamefont {Evans},
  \citenamefont {Ballmer}, \citenamefont {Fejer}, \citenamefont {Fritschel},
  \citenamefont {Harry},\ and\ \citenamefont {Ogin}}]{Evans2008}%
  \BibitemOpen
  \bibfield  {author} {\bibinfo {author} {\bibfnamefont {M.}~\bibnamefont
  {Evans}}, \bibinfo {author} {\bibfnamefont {S.}~\bibnamefont {Ballmer}},
  \bibinfo {author} {\bibfnamefont {M.}~\bibnamefont {Fejer}}, \bibinfo
  {author} {\bibfnamefont {P.}~\bibnamefont {Fritschel}}, \bibinfo {author}
  {\bibfnamefont {G.}~\bibnamefont {Harry}}, \ and\ \bibinfo {author}
  {\bibfnamefont {G.}~\bibnamefont {Ogin}},\ }\href {\doibase
  10.1103/PhysRevD.78.102003} {\bibfield  {journal} {\bibinfo  {journal} {Phys.
  Rev. D}\ }\textbf {\bibinfo {volume} {78}},\ \bibinfo {pages} {102003}
  (\bibinfo {year} {2008})}\BibitemShut {NoStop}%
\bibitem [{\citenamefont {Cerdonio}\ \emph {et~al.}(2001)\citenamefont
  {Cerdonio}, \citenamefont {Conti}, \citenamefont {Heidmann},\ and\
  \citenamefont {Pinard}}]{Cerdonio2001}%
  \BibitemOpen
  \bibfield  {author} {\bibinfo {author} {\bibfnamefont {M.}~\bibnamefont
  {Cerdonio}}, \bibinfo {author} {\bibfnamefont {L.}~\bibnamefont {Conti}},
  \bibinfo {author} {\bibfnamefont {A.}~\bibnamefont {Heidmann}}, \ and\
  \bibinfo {author} {\bibfnamefont {M.}~\bibnamefont {Pinard}},\ }\href
  {\doibase 10.1103/PhysRevD.63.082003} {\bibfield  {journal} {\bibinfo
  {journal} {Phys. Rev. D}\ }\textbf {\bibinfo {volume} {63}},\ \bibinfo
  {pages} {082003} (\bibinfo {year} {2001})}\BibitemShut {NoStop}%
\bibitem [{\citenamefont {Landau}\ and\ \citenamefont {Lifshitz}(1986)}]{L&L}%
  \BibitemOpen
  \bibfield  {author} {\bibinfo {author} {\bibfnamefont {L.~D.}\ \bibnamefont
  {Landau}}\ and\ \bibinfo {author} {\bibfnamefont {E.~M.}\ \bibnamefont
  {Lifshitz}},\ }\href@noop {} {\emph {\bibinfo {title} {Theory of
  Elasticity}}},\ \bibinfo {edition} {3rd}\ ed.\ (\bibinfo  {publisher}
  {Oxford},\ \bibinfo {year} {1986})\BibitemShut {NoStop}%
\bibitem [{\citenamefont {Braginsky}\ \emph {et~al.}(1999)\citenamefont
  {Braginsky}, \citenamefont {Gorodetsky},\ and\ \citenamefont
  {Vyatchanin}}]{BGV1999}%
  \BibitemOpen
  \bibfield  {author} {\bibinfo {author} {\bibfnamefont {V.}~\bibnamefont
  {Braginsky}}, \bibinfo {author} {\bibfnamefont {M.}~\bibnamefont
  {Gorodetsky}}, \ and\ \bibinfo {author} {\bibfnamefont {S.}~\bibnamefont
  {Vyatchanin}},\ }\href {\doibase 10.1016/S0375-9601(99)00785-9} {\bibfield
  {journal} {\bibinfo  {journal} {Physics Letters A}\ }\textbf {\bibinfo
  {volume} {264}},\ \bibinfo {pages} {1 } (\bibinfo {year} {1999})}\BibitemShut
  {NoStop}%
\bibitem [{\citenamefont {Fejer}\ \emph {et~al.}(2004)\citenamefont {Fejer},
  \citenamefont {Rowan}, \citenamefont {Cagnoli}, \citenamefont {Crooks},
  \citenamefont {Gretarsson}, \citenamefont {Harry}, \citenamefont {Hough},
  \citenamefont {Penn}, \citenamefont {Sneddon},\ and\ \citenamefont
  {Vyatchanin}}]{Fejer2004}%
  \BibitemOpen
  \bibfield  {author} {\bibinfo {author} {\bibfnamefont {M.~M.}\ \bibnamefont
  {Fejer}}, \bibinfo {author} {\bibfnamefont {S.}~\bibnamefont {Rowan}},
  \bibinfo {author} {\bibfnamefont {G.}~\bibnamefont {Cagnoli}}, \bibinfo
  {author} {\bibfnamefont {D.~R.~M.}\ \bibnamefont {Crooks}}, \bibinfo {author}
  {\bibfnamefont {A.}~\bibnamefont {Gretarsson}}, \bibinfo {author}
  {\bibfnamefont {G.~M.}\ \bibnamefont {Harry}}, \bibinfo {author}
  {\bibfnamefont {J.}~\bibnamefont {Hough}}, \bibinfo {author} {\bibfnamefont
  {S.~D.}\ \bibnamefont {Penn}}, \bibinfo {author} {\bibfnamefont {P.~H.}\
  \bibnamefont {Sneddon}}, \ and\ \bibinfo {author} {\bibfnamefont {S.~P.}\
  \bibnamefont {Vyatchanin}},\ }\href {\doibase 10.1103/PhysRevD.70.082003}
  {\bibfield  {journal} {\bibinfo  {journal} {Phys. Rev. D}\ }\textbf {\bibinfo
  {volume} {70}},\ \bibinfo {pages} {082003} (\bibinfo {year}
  {2004})}\BibitemShut {NoStop}%
\bibitem [{\citenamefont {Heinert}\ \emph {et~al.}(2011)\citenamefont
  {Heinert}, \citenamefont {Gurkovsky}, \citenamefont {Nawrodt}, \citenamefont
  {Vyatchanin},\ and\ \citenamefont {Yamamoto}}]{Heinert2011}%
  \BibitemOpen
  \bibfield  {author} {\bibinfo {author} {\bibfnamefont {D.}~\bibnamefont
  {Heinert}}, \bibinfo {author} {\bibfnamefont {A.~G.}\ \bibnamefont
  {Gurkovsky}}, \bibinfo {author} {\bibfnamefont {R.}~\bibnamefont {Nawrodt}},
  \bibinfo {author} {\bibfnamefont {S.~P.}\ \bibnamefont {Vyatchanin}}, \ and\
  \bibinfo {author} {\bibfnamefont {K.}~\bibnamefont {Yamamoto}},\ }\href
  {\doibase 10.1103/PhysRevD.84.062001} {\bibfield  {journal} {\bibinfo
  {journal} {Phys. Rev. D}\ }\textbf {\bibinfo {volume} {84}},\ \bibinfo
  {pages} {062001} (\bibinfo {year} {2011})}\BibitemShut {NoStop}%
\bibitem [{\citenamefont {Crooks}\ \emph {et~al.}(2006)\citenamefont {Crooks},
  \citenamefont {Cagnoli}, \citenamefont {Fejer}, \citenamefont {Harry},
  \citenamefont {Hough}, \citenamefont {Khuri-Yakub}, \citenamefont {Penn},
  \citenamefont {Route}, \citenamefont {Rowan}, \citenamefont {Sneddon},
  \citenamefont {Wygant},\ and\ \citenamefont {Yaralioglu}}]{Crooks2006}%
  \BibitemOpen
  \bibfield  {author} {\bibinfo {author} {\bibfnamefont {D.~R.~M.}\
  \bibnamefont {Crooks}}, \bibinfo {author} {\bibfnamefont {G.}~\bibnamefont
  {Cagnoli}}, \bibinfo {author} {\bibfnamefont {M.~M.}\ \bibnamefont {Fejer}},
  \bibinfo {author} {\bibfnamefont {G.}~\bibnamefont {Harry}}, \bibinfo
  {author} {\bibfnamefont {J.}~\bibnamefont {Hough}}, \bibinfo {author}
  {\bibfnamefont {B.~T.}\ \bibnamefont {Khuri-Yakub}}, \bibinfo {author}
  {\bibfnamefont {S.}~\bibnamefont {Penn}}, \bibinfo {author} {\bibfnamefont
  {R.}~\bibnamefont {Route}}, \bibinfo {author} {\bibfnamefont
  {S.}~\bibnamefont {Rowan}}, \bibinfo {author} {\bibfnamefont {P.~H.}\
  \bibnamefont {Sneddon}}, \bibinfo {author} {\bibfnamefont {I.~O.}\
  \bibnamefont {Wygant}}, \ and\ \bibinfo {author} {\bibfnamefont {G.~G.}\
  \bibnamefont {Yaralioglu}},\ }\href
  {http://stacks.iop.org/0264-9381/23/i=15/a=014} {\bibfield  {journal}
  {\bibinfo  {journal} {Classical and Quantum Gravity}\ }\textbf {\bibinfo
  {volume} {23}},\ \bibinfo {pages} {4953} (\bibinfo {year}
  {2006})}\BibitemShut {NoStop}%
\bibitem [{\citenamefont {Bondu}\ \emph {et~al.}(1998)\citenamefont {Bondu},
  \citenamefont {Hello},\ and\ \citenamefont {Vinet}}]{Bondu1998}%
  \BibitemOpen
  \bibfield  {author} {\bibinfo {author} {\bibfnamefont {F.}~\bibnamefont
  {Bondu}}, \bibinfo {author} {\bibfnamefont {P.}~\bibnamefont {Hello}}, \ and\
  \bibinfo {author} {\bibfnamefont {J.-Y.}\ \bibnamefont {Vinet}},\ }\href
  {\doibase 10.1016/S0375-9601(98)00450-2} {\bibfield  {journal} {\bibinfo
  {journal} {Physics Letters A}\ }\textbf {\bibinfo {volume} {246}},\ \bibinfo
  {pages} {227 } (\bibinfo {year} {1998})}\BibitemShut {NoStop}%
\bibitem [{\citenamefont {Braginsky}\ \emph {et~al.}(2000)\citenamefont
  {Braginsky}, \citenamefont {Gorodetsky},\ and\ \citenamefont
  {Vyatchanin}}]{BGV2000}%
  \BibitemOpen
  \bibfield  {author} {\bibinfo {author} {\bibfnamefont {V.}~\bibnamefont
  {Braginsky}}, \bibinfo {author} {\bibfnamefont {M.}~\bibnamefont
  {Gorodetsky}}, \ and\ \bibinfo {author} {\bibfnamefont {S.}~\bibnamefont
  {Vyatchanin}},\ }\href {\doibase 10.1016/S0375-9601(00)00389-3} {\bibfield
  {journal} {\bibinfo  {journal} {Physics Letters A}\ }\textbf {\bibinfo
  {volume} {271}},\ \bibinfo {pages} {303 } (\bibinfo {year}
  {2000})}\BibitemShut {NoStop}%
\bibitem [{\citenamefont {Martin}(2013)}]{Mike2013}%
  \BibitemOpen
  \bibfield  {author} {\bibinfo {author} {\bibfnamefont {M.~J.}\ \bibnamefont
  {Martin}},\ }\emph {\bibinfo {title} {Quantum Metrology and Many-Body
  Physics: Pushing the Frontier of the Optical Lattice Clock}},\ \href@noop {}
  {Ph.D. thesis},\ \bibinfo  {school} {University of Colorado} (\bibinfo {year}
  {2013})\BibitemShut {NoStop}%
\bibitem [{\citenamefont {Kessler}\ \emph
  {et~al.}(2012{\natexlab{a}})\citenamefont {Kessler}, \citenamefont {Legero},\
  and\ \citenamefont {Sterr}}]{Kessler2012a}%
  \BibitemOpen
  \bibfield  {author} {\bibinfo {author} {\bibfnamefont {T.}~\bibnamefont
  {Kessler}}, \bibinfo {author} {\bibfnamefont {T.}~\bibnamefont {Legero}}, \
  and\ \bibinfo {author} {\bibfnamefont {U.}~\bibnamefont {Sterr}},\ }\href
  {\doibase 10.1364/JOSAB.29.000178} {\bibfield  {journal} {\bibinfo  {journal}
  {J. Opt. Soc. Am. B}\ }\textbf {\bibinfo {volume} {29}},\ \bibinfo {pages}
  {178} (\bibinfo {year} {2012}{\natexlab{a}})}\BibitemShut {NoStop}%
\bibitem [{\citenamefont {Numata}\ \emph {et~al.}(2004)\citenamefont {Numata},
  \citenamefont {Kemery},\ and\ \citenamefont {Camp}}]{Numata2004}%
  \BibitemOpen
  \bibfield  {author} {\bibinfo {author} {\bibfnamefont {K.}~\bibnamefont
  {Numata}}, \bibinfo {author} {\bibfnamefont {A.}~\bibnamefont {Kemery}}, \
  and\ \bibinfo {author} {\bibfnamefont {J.}~\bibnamefont {Camp}},\ }\href@noop
  {} {\bibfield  {journal} {\bibinfo  {journal} {Physical Review Letters}\
  }\textbf {\bibinfo {volume} {93}} (\bibinfo {year} {2004})}\BibitemShut
  {NoStop}%
\bibitem [{\citenamefont {Farsi}\ \emph {et~al.}(2012)\citenamefont {Farsi},
  \citenamefont {de~Cumis}, \citenamefont {Marino},\ and\ \citenamefont
  {Marin}}]{Farsi2012}%
  \BibitemOpen
  \bibfield  {author} {\bibinfo {author} {\bibfnamefont {A.}~\bibnamefont
  {Farsi}}, \bibinfo {author} {\bibfnamefont {M.~S.}\ \bibnamefont {de~Cumis}},
  \bibinfo {author} {\bibfnamefont {F.}~\bibnamefont {Marino}}, \ and\ \bibinfo
  {author} {\bibfnamefont {F.}~\bibnamefont {Marin}},\ }\href {\doibase
  10.1063/1.3684626} {\bibfield  {journal} {\bibinfo  {journal} {Journal of
  Applied Physics}\ }\textbf {\bibinfo {volume} {111}},\ \bibinfo {eid}
  {043101} (\bibinfo {year} {2012})}\BibitemShut {NoStop}%
\bibitem [{\citenamefont {Drever}\ \emph {et~al.}(1983)\citenamefont {Drever},
  \citenamefont {Hall}, \citenamefont {Kowalski}, \citenamefont {Hough},
  \citenamefont {Ford}, \citenamefont {Munley},\ and\ \citenamefont
  {Ward}}]{DH1983}%
  \BibitemOpen
  \bibfield  {author} {\bibinfo {author} {\bibfnamefont {R.}~\bibnamefont
  {Drever}}, \bibinfo {author} {\bibfnamefont {J.}~\bibnamefont {Hall}},
  \bibinfo {author} {\bibfnamefont {F.}~\bibnamefont {Kowalski}}, \bibinfo
  {author} {\bibfnamefont {J.}~\bibnamefont {Hough}}, \bibinfo {author}
  {\bibfnamefont {G.}~\bibnamefont {Ford}}, \bibinfo {author} {\bibfnamefont
  {A.}~\bibnamefont {Munley}}, \ and\ \bibinfo {author} {\bibfnamefont
  {H.}~\bibnamefont {Ward}},\ }\href {\doibase 10.1007/BF00702605} {\bibfield
  {journal} {\bibinfo  {journal} {Applied Physics B}\ }\textbf {\bibinfo
  {volume} {31}},\ \bibinfo {pages} {97} (\bibinfo {year} {1983})}\BibitemShut
  {NoStop}%
\bibitem [{\citenamefont {Nazarova}\ \emph {et~al.}(2006)\citenamefont
  {Nazarova}, \citenamefont {Riehle},\ and\ \citenamefont
  {Sterr}}]{Nazarova2006}%
  \BibitemOpen
  \bibfield  {author} {\bibinfo {author} {\bibfnamefont {T.}~\bibnamefont
  {Nazarova}}, \bibinfo {author} {\bibfnamefont {F.}~\bibnamefont {Riehle}}, \
  and\ \bibinfo {author} {\bibfnamefont {U.}~\bibnamefont {Sterr}},\ }\href
  {\doibase 10.1007/s00340-006-2225-y} {\bibfield  {journal} {\bibinfo
  {journal} {Applied Physics B}\ }\textbf {\bibinfo {volume} {83}},\ \bibinfo
  {pages} {531} (\bibinfo {year} {2006})}\BibitemShut {NoStop}%
\bibitem [{\citenamefont {Webster}\ \emph {et~al.}(2007)\citenamefont
  {Webster}, \citenamefont {Oxborrow},\ and\ \citenamefont
  {Gill}}]{Webster2007a}%
  \BibitemOpen
  \bibfield  {author} {\bibinfo {author} {\bibfnamefont {S.~A.}\ \bibnamefont
  {Webster}}, \bibinfo {author} {\bibfnamefont {M.}~\bibnamefont {Oxborrow}}, \
  and\ \bibinfo {author} {\bibfnamefont {P.}~\bibnamefont {Gill}},\ }\href
  {\doibase 10.1103/PhysRevA.75.011801} {\bibfield  {journal} {\bibinfo
  {journal} {Phys. Rev. A}\ }\textbf {\bibinfo {volume} {75}},\ \bibinfo
  {pages} {011801} (\bibinfo {year} {2007})}\BibitemShut {NoStop}%
\bibitem [{\citenamefont {Millo}\ \emph {et~al.}(2009)\citenamefont {Millo},
  \citenamefont {Magalh\~aes}, \citenamefont {Mandache}, \citenamefont
  {Le~Coq}, \citenamefont {English}, \citenamefont {Westergaard}, \citenamefont
  {Lodewyck}, \citenamefont {Bize}, \citenamefont {Lemonde},\ and\
  \citenamefont {Santarelli}}]{Millo2009}%
  \BibitemOpen
  \bibfield  {author} {\bibinfo {author} {\bibfnamefont {J.}~\bibnamefont
  {Millo}}, \bibinfo {author} {\bibfnamefont {D.~V.}\ \bibnamefont
  {Magalh\~aes}}, \bibinfo {author} {\bibfnamefont {C.}~\bibnamefont
  {Mandache}}, \bibinfo {author} {\bibfnamefont {Y.}~\bibnamefont {Le~Coq}},
  \bibinfo {author} {\bibfnamefont {E.~M.~L.}\ \bibnamefont {English}},
  \bibinfo {author} {\bibfnamefont {P.~G.}\ \bibnamefont {Westergaard}},
  \bibinfo {author} {\bibfnamefont {J.}~\bibnamefont {Lodewyck}}, \bibinfo
  {author} {\bibfnamefont {S.}~\bibnamefont {Bize}}, \bibinfo {author}
  {\bibfnamefont {P.}~\bibnamefont {Lemonde}}, \ and\ \bibinfo {author}
  {\bibfnamefont {G.}~\bibnamefont {Santarelli}},\ }\href {\doibase
  10.1103/PhysRevA.79.053829} {\bibfield  {journal} {\bibinfo  {journal} {Phys.
  Rev. A}\ }\textbf {\bibinfo {volume} {79}},\ \bibinfo {pages} {053829}
  (\bibinfo {year} {2009})}\BibitemShut {NoStop}%
\bibitem [{\citenamefont {Fritschel}\ \emph {et~al.}(1998)\citenamefont
  {Fritschel}, \citenamefont {Gonz{\'a}lez}, \citenamefont {Lantz},
  \citenamefont {Saha},\ and\ \citenamefont {Zucker}}]{Fritschel:1998gr}%
  \BibitemOpen
  \bibfield  {author} {\bibinfo {author} {\bibfnamefont {P.}~\bibnamefont
  {Fritschel}}, \bibinfo {author} {\bibfnamefont {G.}~\bibnamefont
  {Gonz{\'a}lez}}, \bibinfo {author} {\bibfnamefont {B.}~\bibnamefont {Lantz}},
  \bibinfo {author} {\bibfnamefont {P.}~\bibnamefont {Saha}}, \ and\ \bibinfo
  {author} {\bibfnamefont {M.}~\bibnamefont {Zucker}},\ }\href@noop {}
  {\bibfield  {journal} {\bibinfo  {journal} {Physical Review Letters}\
  }\textbf {\bibinfo {volume} {80}},\ \bibinfo {pages} {3181} (\bibinfo {year}
  {1998})}\BibitemShut {NoStop}%
\bibitem [{\citenamefont {Adhikari}(2004)}]{Rana2004}%
  \BibitemOpen
  \bibfield  {author} {\bibinfo {author} {\bibfnamefont {R.}~\bibnamefont
  {Adhikari}},\ }\emph {\bibinfo {title} {Sensitivity and Noise Analysis of 4
  km Laser Interferometric Gravitational Wave Antennae}},\ \href@noop {} {Ph.D.
  thesis},\ \bibinfo  {school} {Massachusetts Institute of Technology}
  (\bibinfo {year} {2004})\BibitemShut {NoStop}%
\bibitem [{\citenamefont {Ishibashi}\ \emph {et~al.}(2002)\citenamefont
  {Ishibashi}, \citenamefont {Ye},\ and\ \citenamefont {Hall}}]{Hall2002}%
  \BibitemOpen
  \bibfield  {author} {\bibinfo {author} {\bibfnamefont {C.}~\bibnamefont
  {Ishibashi}}, \bibinfo {author} {\bibfnamefont {J.}~\bibnamefont {Ye}}, \
  and\ \bibinfo {author} {\bibfnamefont {J.}~\bibnamefont {Hall}},\ }in\ \href
  {\doibase 10.1109/QELS.2002.1031144} {\emph {\bibinfo {booktitle} {Quantum
  Electronics and Laser Science Conference, 2002. QELS '02. Technical Digest.
  Summaries of Papers Presented at the}}}\ (\bibinfo {year} {2002})\ pp.\
  \bibinfo {pages} {91--92}\BibitemShut {NoStop}%
\bibitem [{Note1()}]{Note1}%
  \BibitemOpen
  \bibinfo {note} {For silica/tanala QWL coatings, most of the light penetrates
  only into the first few doublets}\BibitemShut {NoStop}%
\bibitem [{\citenamefont {von Toussaint}(2011)}]{vonToussaint2011}%
  \BibitemOpen
  \bibfield  {author} {\bibinfo {author} {\bibfnamefont {U.}~\bibnamefont {von
  Toussaint}},\ }\href {\doibase 10.1103/RevModPhys.83.943} {\bibfield
  {journal} {\bibinfo  {journal} {Rev. Mod. Phys.}\ }\textbf {\bibinfo {volume}
  {83}},\ \bibinfo {pages} {943} (\bibinfo {year} {2011})}\BibitemShut
  {NoStop}%
\bibitem [{Note2()}]{Note2}%
  \BibitemOpen
  \bibinfo {note} {Harry originally determined $\protect \mathaccentV
  {hat}05E{\phi }_\parallel \pm \sigma _{\protect \mathaccentV {hat}05E{\phi
  }_\parallel } = (1.0 \pm 0.3) \times 10^{-4}$ using a coating thickness that
  was 5 times the actual value. Taking into account the correction given in
  Penn et~al.~\cite {Penn2003}, we reanalyze Harry's ringdown data to arrive at
  arrive at $\protect \mathaccentV {hat}05E{\phi }_\parallel \pm \sigma
  _{\protect \mathaccentV {hat}05E{\phi }_\parallel } = (5.2\pm 0.8)\times
  10^{-4}$.}\BibitemShut {Stop}%
\bibitem [{\citenamefont {Cole}\ \emph {et~al.}(2013)\citenamefont {Cole},
  \citenamefont {Zhang}, \citenamefont {Martin}, \citenamefont {Ye},\ and\
  \citenamefont {Aspelmeyer}}]{Cole2013}%
  \BibitemOpen
  \bibfield  {author} {\bibinfo {author} {\bibfnamefont {G.~D.}\ \bibnamefont
  {Cole}}, \bibinfo {author} {\bibfnamefont {W.}~\bibnamefont {Zhang}},
  \bibinfo {author} {\bibfnamefont {M.~J.}\ \bibnamefont {Martin}}, \bibinfo
  {author} {\bibfnamefont {J.}~\bibnamefont {Ye}}, \ and\ \bibinfo {author}
  {\bibfnamefont {M.}~\bibnamefont {Aspelmeyer}},\ }\href {\doibase
  10.1038/nphoton.2013.174} {\bibfield  {journal} {\bibinfo  {journal} {Nature
  Photonics}\ }\textbf {\bibinfo {volume} {7}},\ \bibinfo {pages} {644}
  (\bibinfo {year} {2013})}\BibitemShut {NoStop}%
\bibitem [{\citenamefont {Kessler}\ \emph
  {et~al.}(2012{\natexlab{b}})\citenamefont {Kessler}, \citenamefont
  {Hagemann}, \citenamefont {Grebing}, \citenamefont {Legero}, \citenamefont
  {Sterr}, \citenamefont {Riehle}, \citenamefont {Martin}, \citenamefont
  {Chen},\ and\ \citenamefont {Ye}}]{Kessler2012b}%
  \BibitemOpen
  \bibfield  {author} {\bibinfo {author} {\bibfnamefont {T.}~\bibnamefont
  {Kessler}}, \bibinfo {author} {\bibfnamefont {C.}~\bibnamefont {Hagemann}},
  \bibinfo {author} {\bibfnamefont {C.}~\bibnamefont {Grebing}}, \bibinfo
  {author} {\bibfnamefont {T.}~\bibnamefont {Legero}}, \bibinfo {author}
  {\bibfnamefont {U.}~\bibnamefont {Sterr}}, \bibinfo {author} {\bibfnamefont
  {F.}~\bibnamefont {Riehle}}, \bibinfo {author} {\bibfnamefont
  {M.}~\bibnamefont {Martin}}, \bibinfo {author} {\bibfnamefont
  {L.}~\bibnamefont {Chen}}, \ and\ \bibinfo {author} {\bibfnamefont
  {J.}~\bibnamefont {Ye}},\ }\href@noop {} {\bibfield  {journal} {\bibinfo
  {journal} {Nature Photonics}\ }\textbf {\bibinfo {volume} {6}},\ \bibinfo
  {pages} {687} (\bibinfo {year} {2012}{\natexlab{b}})}\BibitemShut {NoStop}%
\bibitem [{\citenamefont {{Penn}}\ \emph {et~al.}(2003)\citenamefont {{Penn}},
  \citenamefont {{Sneddon}}, \citenamefont {{Armandula}}, \citenamefont
  {{Betzwieser}}, \citenamefont {{Cagnoli}}, \citenamefont {{Camp}},
  \citenamefont {{Crooks}}, \citenamefont {{Fejer}}, \citenamefont
  {{Gretarsson}}, \citenamefont {{Harry}}, \citenamefont {{Hough}},
  \citenamefont {{Kittelberger}}, \citenamefont {{Mortonson}}, \citenamefont
  {{Route}}, \citenamefont {{Rowan}},\ and\ \citenamefont
  {{Vassiliou}}}]{Penn2003}%
  \BibitemOpen
  \bibfield  {author} {\bibinfo {author} {\bibfnamefont {S.~D.}\ \bibnamefont
  {{Penn}}}, \bibinfo {author} {\bibfnamefont {P.~H.}\ \bibnamefont
  {{Sneddon}}}, \bibinfo {author} {\bibfnamefont {H.}~\bibnamefont
  {{Armandula}}}, \bibinfo {author} {\bibfnamefont {J.~C.}\ \bibnamefont
  {{Betzwieser}}}, \bibinfo {author} {\bibfnamefont {G.}~\bibnamefont
  {{Cagnoli}}}, \bibinfo {author} {\bibfnamefont {J.}~\bibnamefont {{Camp}}},
  \bibinfo {author} {\bibfnamefont {D.~R.~M.}\ \bibnamefont {{Crooks}}},
  \bibinfo {author} {\bibfnamefont {M.~M.}\ \bibnamefont {{Fejer}}}, \bibinfo
  {author} {\bibfnamefont {A.~M.}\ \bibnamefont {{Gretarsson}}}, \bibinfo
  {author} {\bibfnamefont {G.~M.}\ \bibnamefont {{Harry}}}, \bibinfo {author}
  {\bibfnamefont {J.}~\bibnamefont {{Hough}}}, \bibinfo {author} {\bibfnamefont
  {S.~E.}\ \bibnamefont {{Kittelberger}}}, \bibinfo {author} {\bibfnamefont
  {M.~J.}\ \bibnamefont {{Mortonson}}}, \bibinfo {author} {\bibfnamefont
  {R.}~\bibnamefont {{Route}}}, \bibinfo {author} {\bibfnamefont
  {S.}~\bibnamefont {{Rowan}}}, \ and\ \bibinfo {author} {\bibfnamefont
  {C.~C.}\ \bibnamefont {{Vassiliou}}},\ }\href {\doibase
  10.1088/0264-9381/20/13/334} {\bibfield  {journal} {\bibinfo  {journal}
  {Classical and Quantum Gravity}\ }\textbf {\bibinfo {volume} {20}},\ \bibinfo
  {pages} {2917} (\bibinfo {year} {2003})},\ \Eprint
  {http://arxiv.org/abs/arXiv:gr-qc/0302093} {arXiv:gr-qc/0302093} \BibitemShut
  {NoStop}%
\bibitem [{\citenamefont {Crooks}\ \emph {et~al.}(2004)\citenamefont {Crooks},
  \citenamefont {Cagnoli}, \citenamefont {Fejer}, \citenamefont {Gretarsson},
  \citenamefont {Harry}, \citenamefont {Hough}, \citenamefont {Nakagawa},
  \citenamefont {Penn}, \citenamefont {Route}, \citenamefont {Rowan},\ and\
  \citenamefont {Sneddon}}]{Crooks2004}%
  \BibitemOpen
  \bibfield  {author} {\bibinfo {author} {\bibfnamefont {D.~R.~M.}\
  \bibnamefont {Crooks}}, \bibinfo {author} {\bibfnamefont {G.}~\bibnamefont
  {Cagnoli}}, \bibinfo {author} {\bibfnamefont {M.~M.}\ \bibnamefont {Fejer}},
  \bibinfo {author} {\bibfnamefont {A.}~\bibnamefont {Gretarsson}}, \bibinfo
  {author} {\bibfnamefont {G.}~\bibnamefont {Harry}}, \bibinfo {author}
  {\bibfnamefont {J.}~\bibnamefont {Hough}}, \bibinfo {author} {\bibfnamefont
  {N.}~\bibnamefont {Nakagawa}}, \bibinfo {author} {\bibfnamefont
  {S.}~\bibnamefont {Penn}}, \bibinfo {author} {\bibfnamefont {R.}~\bibnamefont
  {Route}}, \bibinfo {author} {\bibfnamefont {S.}~\bibnamefont {Rowan}}, \ and\
  \bibinfo {author} {\bibfnamefont {P.~H.}\ \bibnamefont {Sneddon}},\ }\href
  {http://stacks.iop.org/0264-9381/21/i=5/a=101} {\bibfield  {journal}
  {\bibinfo  {journal} {Classical and Quantum Gravity}\ }\textbf {\bibinfo
  {volume} {21}},\ \bibinfo {pages} {S1059} (\bibinfo {year}
  {2004})}\BibitemShut {NoStop}%
\bibitem [{\citenamefont {{Li}}\ \emph {et~al.}(2014)\citenamefont {{Li}},
  \citenamefont {{Aguilar Sandoval}}, \citenamefont {{Geitner}}, \citenamefont
  {{Cagnoli}}, \citenamefont {{Dolique}}, \citenamefont {{Degallaix}},
  \citenamefont {{Flaminio}}, \citenamefont {{Forest}}, \citenamefont
  {{Granata}}, \citenamefont {{Michel}}, \citenamefont {{Morgado}},
  \citenamefont {{Pinard}},\ and\ \citenamefont {{Bellon}}}]{Li2014}%
  \BibitemOpen
  \bibfield  {author} {\bibinfo {author} {\bibfnamefont {T.}~\bibnamefont
  {{Li}}}, \bibinfo {author} {\bibfnamefont {F.~A.}\ \bibnamefont {{Aguilar
  Sandoval}}}, \bibinfo {author} {\bibfnamefont {M.}~\bibnamefont {{Geitner}}},
  \bibinfo {author} {\bibfnamefont {G.}~\bibnamefont {{Cagnoli}}}, \bibinfo
  {author} {\bibfnamefont {V.}~\bibnamefont {{Dolique}}}, \bibinfo {author}
  {\bibfnamefont {J.}~\bibnamefont {{Degallaix}}}, \bibinfo {author}
  {\bibfnamefont {R.}~\bibnamefont {{Flaminio}}}, \bibinfo {author}
  {\bibfnamefont {D.}~\bibnamefont {{Forest}}}, \bibinfo {author}
  {\bibfnamefont {M.}~\bibnamefont {{Granata}}}, \bibinfo {author}
  {\bibfnamefont {C.}~\bibnamefont {{Michel}}}, \bibinfo {author}
  {\bibfnamefont {N.}~\bibnamefont {{Morgado}}}, \bibinfo {author}
  {\bibfnamefont {L.}~\bibnamefont {{Pinard}}}, \ and\ \bibinfo {author}
  {\bibfnamefont {L.}~\bibnamefont {{Bellon}}},\ }\href@noop {} {\bibfield
  {journal} {\bibinfo  {journal} {ArXiv e-prints}\ } (\bibinfo {year}
  {2014})},\ \Eprint {http://arxiv.org/abs/1401.0184} {arXiv:1401.0184}
  \BibitemShut {NoStop}%
\bibitem [{\citenamefont {{Johannes Eichholz and Michael
  Hartman}}(2013)}]{Eichholz2013}%
  \BibitemOpen
  \bibfield  {author} {\bibinfo {author} {\bibnamefont {{Johannes Eichholz and
  Michael Hartman}}}\ }(\bibinfo  {publisher} {LVC meeting, Hannover,
  Germany},\ \bibinfo {year} {2013})\BibitemShut {NoStop}%
\bibitem [{\citenamefont {Hartman}\ \emph {et~al.}(2014)\citenamefont
  {Hartman}, \citenamefont {Eichholz}, \citenamefont {Fulda}, \citenamefont
  {Ciani}, \citenamefont {Tanner},\ and\ \citenamefont
  {Mueller}}]{hartman2014measurement}%
  \BibitemOpen
  \bibfield  {author} {\bibinfo {author} {\bibfnamefont {M.}~\bibnamefont
  {Hartman}}, \bibinfo {author} {\bibfnamefont {J.}~\bibnamefont {Eichholz}},
  \bibinfo {author} {\bibfnamefont {P.}~\bibnamefont {Fulda}}, \bibinfo
  {author} {\bibfnamefont {G.}~\bibnamefont {Ciani}}, \bibinfo {author}
  {\bibfnamefont {D.~B.}\ \bibnamefont {Tanner}}, \ and\ \bibinfo {author}
  {\bibfnamefont {G.}~\bibnamefont {Mueller}},\ }\href@noop {} {\bibfield
  {journal} {\bibinfo  {journal} {Bulletin of the American Physical Society}\
  }\textbf {\bibinfo {volume} {59}} (\bibinfo {year} {2014})}\BibitemShut
  {NoStop}%
\bibitem [{\citenamefont {Adhikari}(2014)}]{Rana:RMP2014}%
  \BibitemOpen
  \bibfield  {author} {\bibinfo {author} {\bibfnamefont {R.~X.}\ \bibnamefont
  {Adhikari}},\ }\href {\doibase 10.1103/RevModPhys.86.121} {\bibfield
  {journal} {\bibinfo  {journal} {Rev. Mod. Phys.}\ }\textbf {\bibinfo {volume}
  {86}},\ \bibinfo {pages} {121} (\bibinfo {year} {2014})}\BibitemShut
  {NoStop}%
\bibitem [{\citenamefont {Tara~Chalermsongsak}(prep)}]{AlGaAs:TO}%
  \BibitemOpen
  \bibfield  {author} {\bibinfo {author} {\bibfnamefont {R.~X.~A.}\
  \bibnamefont {Tara~Chalermsongsak}, \bibfnamefont {Garrett D.~Cole}},\
  }\href@noop {} {\  (\bibinfo {year} {\textit{in prep.}})}\BibitemShut
  {NoStop}%
\bibitem [{\citenamefont {Bondu}\ \emph {et~al.}(1996)\citenamefont {Bondu},
  \citenamefont {Fritschel}, \citenamefont {Man},\ and\ \citenamefont
  {Brillet}}]{Bondu1996}%
  \BibitemOpen
  \bibfield  {author} {\bibinfo {author} {\bibfnamefont {F.}~\bibnamefont
  {Bondu}}, \bibinfo {author} {\bibfnamefont {P.}~\bibnamefont {Fritschel}},
  \bibinfo {author} {\bibfnamefont {C.~N.}\ \bibnamefont {Man}}, \ and\
  \bibinfo {author} {\bibfnamefont {A.}~\bibnamefont {Brillet}},\ }\href
  {\doibase 10.1364/OL.21.000582} {\bibfield  {journal} {\bibinfo  {journal}
  {Optics Letters}\ }\textbf {\bibinfo {volume} {21}},\ \bibinfo {pages} {582}
  (\bibinfo {year} {1996})}\BibitemShut {NoStop}%
\end{thebibliography}%

\end{document}